\documentclass[final,twocolumn]{svjour3}%
\usepackage{graphicx}
\usepackage{subfigure}
\usepackage{CJK}
\usepackage{amsmath}
\usepackage{amssymb}
\usepackage{mathptmx}
\usepackage{latexsym}
\usepackage{verbatim}
\usepackage{color}
\usepackage[english]{babel}
\usepackage[breaklinks=true,colorlinks=true,linkcolor=blue,citecolor=blue,urlcolor=black,bookmarks=false]%
{hyperref}
\usepackage{amsfonts}%
\providecommand{\U}[1]{\protect\rule{.1in}{.1in}}
\RequirePackage{fix-cm}
\journalname{Journal of Computational Electronics}
\begin{document}
\begin{CJK}{Bg5}{bsmi}
\title{Gate-induced carrier density modulation in bulk graphene: Theories and
electrostatic simulation using \textsc{Matlab} pdetool}
\author{Ming-Hao Liu ({\rm ¼B©ú»¨})}
\institute{Ming-Hao Liu (¼B©ú»¨) \at\ \mbox{Institut f\"{u}r Theoretische Physik,
Universit\"{a}t Regensburg,} \mbox{D-93040 Regensburg, Germany} \at \mbox{\email{minghao.liu.taiwan@gmail.com}}}
\date{Received: date / Accepted: date}
\maketitle
\sloppy
\begin{abstract}
This article aims at providing a self-contained introduction to
theoretical modeling of gate-induced carrier density in graphene
sheets. For this, relevant theories are introduced, namely,
classical capacitance model (CCM), self-consistent Poisson-Dirac
method (PDM), and quantum capacitance model (QCM). The usage of
\textsc{Matlab} pdetool is also briefly introduced, pointing out the
least knowledge required for using this tool to solve the present
electrostatic problem. Results based on the three approaches are
compared, showing that the quantum correction, which is not
considered by the CCM but by the other two, plays a role only when
the metal gate is exceedingly close to the graphene sheet, and that
the exactly solvable QCM works equally well as the self-consistent
PDM. Practical examples corresponding to realistic experimental
conditions for generating graphene \textit{pnp} junctions and
superlattices, as well as how a background potential linear in
position can be achieved in graphene, are shown to illustrate the
applicability of the introduced methods. Furthermore, by treating
metal contacts in the same way, the last example shows that the PDM
and the QCM are able to resolve the contact-induced doping and
screening potential, well agreeing with the previous
first-principles studies. 
%\subclass{MSC code1 \and MSC code2 \and more}
\end{abstract}
\end{CJK}

\sloppy

\section{Introduction}

Electronic transport in graphene \cite{CastroNeto2009,DasSarma2011}, a
one-atom-thick honeycomb carbon lattice, is one of its main issues among the
increasing number of fundamental studies ever since the first successful
isolation of stable monolayer graphene flakes in 2004 \cite{Novoselov2004}.
What led to the explosive growth of the graphene literature, however, was not
only the discovery of the mechanical exfoliation (Scotch-tape method) for
graphene flake preparation, which made graphene easily accessible to
laboratories all over the world, but also the characterization of the
electronic properties of graphene by electrical gating, which provides a
direct way to modulate the carrier density, and hence the Fermi level, of
graphene \cite{Novoselov2004}. Conductance (resistance) sweep using a single
backgate is henceforth a standard electronic characterization tool for
graphene. Double-gated graphene opens even more possibilities of graphene
electronics and allows experimental studies of graphene \emph{pn} and
\emph{pnp} junctions
\cite{Huard2007,Williams2007,Ozyilmaz2007,Liu2008c,Gorbachev2008}, as well as
the interesting physics of Klein tunneling
\cite{Cheianov2006,Katsnelson2006,Stander2009,Young2009,Nam2011}. Gate-induced
carrier density modulation, therefore, plays an essential role for fundamental
as well as advanced studies of graphene electronics.

Theory of the gate-induced carrier density modulation is mainly an
electrostatic problem. How one should obtain the gate-voltage dependence of
the carrier density in graphene depends actually on how precise one wishes.
For cheapest computation, the graphene sheet carrier density can be directly
regarded as the induced surface charge density adjacent to graphene
\cite{Novoselov2004}, which is treated as a conductor fixed at zero potential.
This corresponds to the classical capacitance model (CCM) that is widely
adopted in most experimental works on graphene transport \cite{DasSarma2011}
and can be solved exactly. A more precise computation takes into account the
relation between the induced charge density on graphene and the electric
potential energy that those charge carriers gain, through the graphene density
of states \cite{Guo2007,Fern'andez-Rossier2007,Fang2007}. This requires
self-consistent iterative computation
\cite{Gorbachev2008,Shylau2009,Andrijauskas2012}, which is a bit more
expensive, but actually corresponds to the quantum capacitance model (QCM)
\cite{Luryi1988}, within which exact solutions for single-gated pristine
graphene \cite{Fang2007} and even multigated doped graphene \cite{Liu2013} can
be obtained. Further considerations such as the Coulomb interaction of the
induced charges on the graphene sheet are possible
\cite{Fern'andez-Rossier2007,Shylau2009}, but these would be out of the scope
of the present discussion.

Whereas a thorough and comprehensive review on the theory of gate-induced
carrier density modulation of bulk graphene so far does not exist in the
literature, part of this article aims at providing this missing piece. The
review includes both the analytical and numerical aspects, as well as a brief
introduction to the usage of \textsc{Matlab}'s pdetool, in order for a
self-contained context. Readers who happen to be \textsc{Matlab} users would
find this brief usage helpful, but non-\textsc{Matlab} users may as well
neglect it without encountering further gaps. The analytics based on the CCM
and QCM and the numerics based on the self-consistent iteration method,
namely, the Poisson-Dirac method (PDM), using \textsc{Matlab}'s pdetool will
be compared, showing that the quantum correction plays usually a minor role,
unless the metal gate is exceedingly close to the graphene sheet. In the case
of single-gated pristine graphene, consistency between the QCM and PDM is
satisfactory even for capacitors with finite gates, and is exact for
parallel-plate capacitors with infinitely extending gates.

With a full understanding of the gate-voltage modulation on the graphene
carrier density, examples of its applications aiming at providing realistic
local energy band offsets due to electric gating will be illustrated. This is
particularly important for an accurate theoretical modeling of transport in
graphene \cite{Liu2012a}. Examples include (i) graphene \emph{pnp} junctions,
(ii) graphene superlattices, and (iii) generation of background potential
linear in position in graphene. Practically, the example (i) provides the
study of the physics of Klein backscattering
\cite{Shytov2008,Young2009,Liu2012a}, while the combination of the examples
(ii) and (iii) is the underlying prerequisite of the Bloch-Zener oscillation
in graphene \cite{Krueckl2012}. Furthermore, the introduced PDM and QCM are
capable of treating the effects of metal contacts, for which example (iv) of
contact-induced doping and screening potential is also illustrated. Taking
palladium as a specific example of the metal contact, the results obtained by
the PDM and QCM are shown to agree well with the previous first-principles
studies \cite{Khomyakov2009,Khomyakov2010}.

This paper is organized as follows. In Sec.\ \ref{sec 2}, we first provide a
brief introduction to the usage of \textsc{Matlab}'s pdetool, pointing out the
least knowledge required to apply the tool on the present specific
electrostatic problem. Theories of the gate-induced carrier density modulation
in graphene are reviewed in Sec.\ \ref{sec 3}, where the analytics based on
the capacitance models and the numerics based on the PDM is also compared.
Practical applications based on the theories reviewed in Sec.\ \ref{sec 3} are
given in Sec.\ \ref{sec 4}, and a summary of the present work is concluded in
Sec.\ \ref{sec 5}.

\section{Usage of Matlab's pdetool for electrostatics\label{sec 2}}

The pdetool is a useful numerical tool built in \textsc{Matlab} and provides a
convenient way to solve several classic partial differential equation (PDE)
problems in two-dimension. For the electrostatics at our present interest, the
Poisson equation,
\begin{equation}
-\vec{\nabla}\cdot(\epsilon_{r}\vec{\nabla}u)=\frac{\rho}{\epsilon_{0}},
\label{poisson eq}%
\end{equation}
obtained from two of the Maxwell's equations, $\vec{\nabla}\times\mathbf{E}=0$
and $\vec{\nabla}\cdot\mathbf{D}=\rho$, which respectively lead to
$\mathbf{E}=-\vec{\nabla}u$ and $\vec{\nabla}\cdot\epsilon_{r}\mathbf{E}%
=\rho/\epsilon_{0}$, is the central equation that the pdetool solves for the
electric potential $u$.\footnote{To be consistent with the pdetool, we name
the electric potential as $u$, while reserve the variable $V$ for the energy
band offset (the \textquotedblleft on-site energy\textquotedblright\ in the
language of tight-binding formulation).} In Eq.\ \eqref{poisson eq}, the
product of the dielectric constant (relative permittivity) $\epsilon_{r}$ and
the free space permittivity $\epsilon_{0}$ gives the absolute permittivity
$\epsilon=\epsilon_{r}\epsilon_{0}$.

A full introduction to the usage of the pdetool can be found in the
\textsc{Matlab} documentation \cite{pde} and need not be repeated here. To
digest the full user's guide of the tool, however, is not necessary for our
present focus, which is essentially an electrostatic problem. This section is
basically to elaborate those that are less clear in \cite{pde} but
nevertheless important for our purpose of obtaining the gate-voltage
dependence of the graphene carrier density, and to point out the least
required knowledge for this purpose.

\subsection{Overview of pdetool}

To solve a PDE problem using the pdetool, required necessary inputs can be
exported from the graphical user interface (GUI) of the pdetool (initiated by
executing \textquotedblleft pdetool\textquotedblright\ from the command
window) and are briefly described in the following.\renewcommand{\labelenumi}{(\roman{enumi})}

\begin{enumerate}
\item \emph{System geometry}. The geometrical shapes of the building blocks,
such as the oxide layers, metallic gates, etc., which constitute the system
where the PDE problem is defined, can be drawn in the \textquotedblleft Draw
Mode\textquotedblright\ of the GUI. The resulted \textquotedblleft decomposed
geometry\textquotedblright\ allows us to proceed to the rest of the inputs,
but there is no need to \textquotedblleft Export Decomposed Geometry, Set
Formula, Labels...\textquotedblright\ from the \textquotedblleft Draw
menu\textquotedblright\ since not all of them will be needed by the PDE solvers.

\item \emph{PDE coefficients}. In the \textquotedblleft PDE
Mode\textquotedblright\ of the GUI, one can designate different regions of
materials by filling in the respective dielectric constants and space charge
densities. These are stored in certain PDE coefficients matrices, which can be
output from the GUI and will be required by the PDE solvers in programming.

\item \emph{Boundary conditions}. In the \textquotedblleft Boundary
Mode\textquotedblright\ of the GUI, boundary conditions for each bounding edge
can be assigned. The resulting boundary matrix $b$, which will be required by
the PDE solvers in programming, and the Decomposed Geometry, which will be
required when visualizing the PDE geometry, can be exported from the
\textquotedblleft Boundary menu\textquotedblright. An elaborated instruction
about $b$ will be given later.

\item \emph{Mesh points}. The mesh points are those spatial points at which
the numerical solutions are desired.\ They can be created, refined, or jiggled
in the \textquotedblleft Mesh Mode\textquotedblright\ of the GUI. The
resulting triangular mesh data, stored by point, edge, and triangle matrices,
can be exported by the GUI and will be used not only when calling for the PDE
solvers but also when visualizing the solution.
\end{enumerate}

With all these requirements completed, the PDE problem is then defined, and
the solution can as well be obtained by clicking \textquotedblleft Solve
PDE\textquotedblright\ within the GUI, which is user-friendly but cannot be
\textquotedblleft programmed\textquotedblright. When performing certain real
calculations, however, especially when a systematic change of variables is
required, programming with, e.g., looping, is inevitable and the requirements
of (ii)--(iv) will be the necessary inputs of the PDE solvers. For our purpose
of simulating the carrier density modulation due to gating, we would often
need to change the gate voltages, which are described by the boundary
conditions. Thus although each of (ii)--(iv) can be programmed by using
relevant commands, in the following only the implementation of (iii) by
commands will be described in detail.

\subsection{Boundary conditions}

\subsubsection{The boundary condition matrix: General
description\label{sec boundary condition matrix}}

By searching \textquotedblleft assemb\textquotedblright\ from the
\textsc{Matlab} help or by looking up in its documentation \cite{pde}, we see
that the boundary conditions are saved in a matrix called $b$, with the
following data format:\renewcommand{\labelitemi}{\textbullet}

\begin{itemize}
\item Row $1$ contains the dimension $N$ of the system. (Note: normally $N=1$.
If one solves two coupled variables, then $N=2,$ etc.; by examining the
exported boundary condition matrix $b$, one would find that $N=0$ for inner boundaries.)

\item Row $2$ contains the number $M$ of Dirichlet boundary conditions.

\item Rows $3$ to $3+N^{2}-1$ contain the lengths for the strings representing
$q$. The lengths are stored in columnwise order with respect to $q$. [See
Eq.\ \eqref{Neumann BC} below.]

\item Rows $3+N^{2}$ to $3+N^{2}+N-1$ contain the lengths for the strings
representing $g$. [See Eq.\ \eqref{Neumann BC} below.]

\item Rows $3+N^{2}+N$ to $3+N^{2}+N+MN-1$ contain the lengths for the strings
representing $h$. The lengths are stored in columnwise order with respect to
$h$. [See Eq.\ \eqref{Dirichlet BC} below.]

\item Rows $3+N^{2}+N+NM$ to $3+N^{2}+N+MN+M-1$ contain the lengths for the
strings representing $r$. [See Eq.\ \eqref{Dirichlet BC} below.]

\item The following rows contain \emph{text expressions} representing the
actual boundary condition functions.
\end{itemize}

Here, two types of boundary conditions\footnote{The mixed type boundary
conditions will not be encountered in the present discussion.} are included,
namely, the Neumann boundary%
\begin{equation}
\mathbf{n}\cdot(c\vec{\nabla}u)+qu=g, \label{Neumann BC}%
\end{equation}
and the Dirichlet boundary%
\begin{equation}
hu=r. \label{Dirichlet BC}%
\end{equation}
In Eq.\ \eqref{Neumann BC}, $c$ contains the PDE coefficients (here the
dielectric constants in different regions), and $\mathbf{n}$ is the normal of
the boundary. So the boundary conditions for given gate voltages would be the
Dirichlet type, with $h=1$ and $r$ being the corresponding voltage. For the
Neumann type boundary condition, we normally consider $q=0$, and $g$
represents the surface charge.

In the following, let us be more specific about the format of the boundary
conditions matrix, considering the two types of boundaries with $N=1$.

\subsubsection{Dirichlet boundary}

Following the general description of Sec.\ \ref{sec boundary condition matrix}%
, the boundary matrix $b$ for a Dirichlet boundary is described as follows.

\begin{itemize}
\item Row $1$ contains the dimension $N$ of the system: $1.$

\item Row $2$ contains the number $M$ of Dirichlet boundary conditions: $1.$

\item Row $3$ contains the length for the strings representing $q$, which is
$1$ since $q=0$, though not used.

\item Row $4$ contains the length for the strings representing $g$, which is
$1$ since $g=0$, though not used.

\item Row $5$ contains the length for the strings representing $h$, which is
$1$ since $h=1$.

\item Row $6$ contains the length for the strings representing $r$.

\item Then comes the text expressions of $q,g,h,r$.
\end{itemize}

An example of a Dirichlet boundary with $r=3.5$ would be:%
\[
\text{\texttt{b = [1 1 1 1 1 3 '0' '0' '1' '3.5']';}}%
\]
The boundary condition may include the $x$ and $y$ position coordinates and
their functions, and can be written even in terms of the solution $u$
(nonlinear solver required). For example,%
\[
\text{\texttt{b = [1 1 1 1 1 4 '0' '0' '1' 'x.\symbol{94}2']';}}%
\]
For another example,
\[
\text{\texttt{b = [1 1 1 1 1 9 '0' '0' '1' 'sin(x).\symbol{94}u']';}}%
\]

\subsubsection{Neumann boundary}

Following the general description of Sec.\ \ref{sec boundary condition matrix}%
, the boundary matrix $b$ for a Neumann boundary is described as follows.

\begin{itemize}
\item Row $1$ contains the dimension $N$ of the system: $1$.

\item Row $2$ contains the number $M$ of Dirichlet boundary conditions: $0$.

\item Row $3$ contains the length for the strings representing $q$, which is
$1$ since $q=0$.

\item Row $4$ contains the length for the strings representing $g$.

\item Then comes the text expressions of $q,g$.
\end{itemize}

An example of a Neumann boundary with surface charge $g=1.6$ would be%
\[
\text{\texttt{b = [1 0 1 3 '0' '1.6']';}}%
\]
Another example%
\[
\text{\texttt{b = [1 0 1 21 '0' '-13.295*sign(u).*u.\symbol{94}2']';}}%
\]
will actually be used when we apply the Poisson-Dirac iteration method.

\subsubsection{Text expression of the boundary conditions}

The boundary condition matrix $b$ exported from the GUI of the pdetool looks
filled with purely integers. This is the \textquotedblleft number
representation\textquotedblright\ of the text strings. For example, a number
$48$ within the $b$ matrix actually means `0':
\begin{verbatim}
>> char(48)

ans =

0
\end{verbatim}

%

%TCIMACRO{\TeXButton{no indent}{\noindent}}%
%BeginExpansion
\noindent
%EndExpansion
Conversely, if we want to transform the strings into numbers, we can simply
use the `\texttt{double}' command:
\begin{verbatim}
>> double('x.^2')

ans =

   120    46    94    50
\end{verbatim}

%

%TCIMACRO{\TeXButton{no indent}{\noindent}}%
%BeginExpansion
\noindent
%EndExpansion
Thus to enter a boundary condition of, e.g., $-(x^{2}+y^{2})$, we may fill in
with:%
\[
\text{\texttt{double('-(x.\symbol{94}2+y.\symbol{94}2)')'}}%
\]
To enter a Dirichlet boundary condition of a given number assigned by a
variable named \texttt{Vtg}, we may fill in with:%
\[
\text{\texttt{double(num2str(Vtg))'}}%
\]
Note that the operator \texttt{'} at the end of these two examples is to take
transpose of the converted text strings, since the boundary conditions are
saved columnwise in $b$, and similarly in the previous examples for $b$.

For a real PDE problem, the number of columns of the $b$ matrix depends on the
total number of edges, including inner and outer boundaries. The $n$th column
records the boundary condition for the $n$th edge. Thus before exporting the
initial $b$ matrix from the GUI of pdetool, one has to check the boundary
labels corresponding to, e.g., graphene or gates (by showing the edge labels
in the \textquotedblleft Boundary Mode\textquotedblright).

\subsection{Some important commands\label{sec commands}}

\subsubsection{Solving the PDE}

A standard PDE solver is called \texttt{assempde}. An example for its usage is
as follows.
\begin{verbatim}
u = assempde(b,p,e,t,c,a,f);
% b: matrix of boundary conditions
% p: points of the mesh grid
% e: edges
% t: triangles
% c,a,f: coefficients of the pde problem
\end{verbatim}

When the solution is involved in the boundary conditions, the solution mode
has to be switched to nonlinear. An example for its usage is as follows.
\begin{verbatim}
[u,res] = pdenonlin(b,p,e,t,c,a,f, ...
    'report','on','MaxIter',1e5,'u0',u0);
% u: the solution, res: not important here
% b,p,e,t,c,a,f: same as above
% 'report':iteration process report
% 'MaxIter': maximal number of iter. rounds
% 'u0': initial guess of the solution
\end{verbatim}

\subsubsection{Interpolation}

To find the values at those points one desires, an important command called
\texttt{tri2grid} should be used, which interpolates from the PDE triangular
mesh to a given rectangular grid. An example for usage is as follows.
\begin{verbatim}
uxy = tri2grid(p,t,u,x,y);
% u: the obtained solution
% p,t: same as above
% x,y: rectangular grid points for interpol.
\end{verbatim}

\subsection{Remarks on units}

In the pdetool, everything is displayed with dimensionless numbers. The actual
units can be chosen as what we would like. Deducing relevant coefficients for
a specific set of chosen units is therefore important before we use the
pdetool to solve any actual problems.

In addition to the physical units, the free space permittivity $\epsilon_{0}$
is suppressed throughout the program. Recall the boundary conditions for the
displacement field $D$ at a conductor-dielectric boundary, which can be
derived by applying the Gauss's law: $D_{t}=0$ and $D_{n}=\rho_{s}$, where
$D_{t}$ and $D_{n}$ represent the components tangential and normal to the
interface, respectively. The normal component of the displacement field
$D_{n}$ therefore means the surface charge density:%
\begin{equation}
-\epsilon_{r}\vec{\nabla}u\cdot\mathbf{n}=\frac{\rho_{s}}{\epsilon_{0}}.
\label{surface charge density}%
\end{equation}
Comparing Eq.\ \eqref{surface charge density} to Eq.\ \eqref{Neumann BC} with
$q=0$ and $c$ representing $\epsilon_{r}$ (instead of $\epsilon$), one can see
that the \textquotedblleft Surface charge\textquotedblright\ actually means%
\begin{equation}
g=\frac{\rho_{s}}{\epsilon_{0}}, \label{g}%
\end{equation}
when filling the Neumann boundary condition in the \textquotedblleft Boundary
Mode\textquotedblright\ of the GUI. Similarly, when filling the PDE
coefficients in the \textquotedblleft PDE Mode\textquotedblright, the
\textquotedblleft Space charge density\textquotedblright\ actually means
\texttt{rho }$=\rho/\epsilon_{0}$, i.e., the right-hand side of Eq.\ \eqref{poisson eq}.

\section{Theories of gate-induced carrier density modulation\label{sec 3}}

In this section, analytical theories of the gate-induced carrier density
modulation, including the classical and quantum capacitance models, are
briefly reviewed, a numerical scheme of the self-consistent Poisson-Dirac
iteration method is introduced, and a numerical comparison between analytics
and numerics is provided.

\subsection{Classical capacitance model}

\subsubsection{The model}

\begin{figure}[t]
\subfigure[]{
\includegraphics[width=\columnwidth]{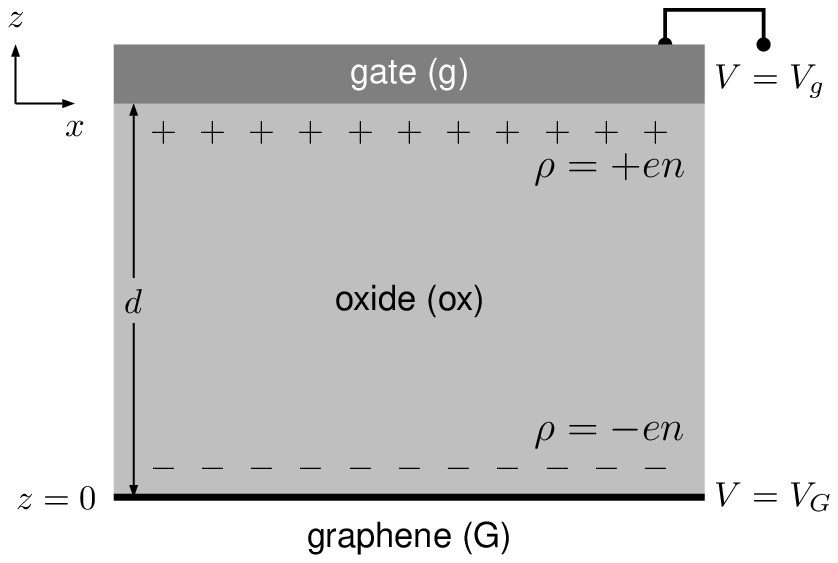}
\label{fig1a}}
\par
\centering
\subfigure[]{
\includegraphics{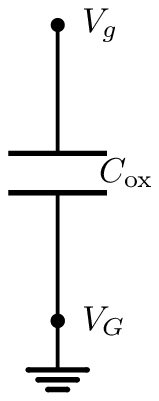}
\label{fig1b}} \subfigure[]{
\includegraphics{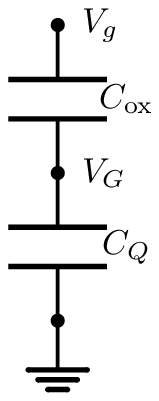}
\label{fig1c}}\caption{(a) Schematic of a single-gated graphene. (b)
Equivalent circuit plot of the classical capacitance model. (c) Equivalent
circuit plot of the quantum capacitance model.}%
\end{figure}

We begin with the classical capacitance model, which considers a
parallel-plate capacitor composed of an oxide dielectric with permittivity
$\epsilon=\epsilon_{r}\epsilon_{0}$ sandwiched by a metallic gate (at $z=d$)
and a conducting graphene sheet (at $z=0$) as sketched in Fig.\ \ref{fig1a}.
Let the electric potential at the gate be $u(x,z=d)=V_{g}$ and the graphene
layer be grounded: $u(x,z=0)=V_{G}=0$. The surface charge density at $z=0^{+}$
(the surface of the oxide dielectric in contact with the graphene layer) from
Eq.\ \eqref{surface charge density} is given by%
\begin{equation}
\rho_{s}=-\epsilon\left.  \frac{\partial u}{\partial z}\right\vert _{z=0^{+}%
}=-\epsilon\frac{V_{g}-0}{d-0}=-C_{\text{ox}}V_{g}, \label{rhos in CM}%
\end{equation}
where $C_{\text{ox}}=\epsilon/d$ is the classical capacitance (per unit area)
of a uniform parallel-plate capacitor. Regarding this surface charge
\eqref{rhos in CM} directly as those induced carriers in the graphene layer,
we have the carrier density%
\begin{equation}
n_{C}=\frac{\rho_{s}}{-e}=\frac{C_{\text{ox}}}{e}V_{g}, \label{nC}%
\end{equation}
which is a widely used formula for estimating the graphene carrier density
\cite{DasSarma2011}. For uniform capacitors with $C_{\text{ox}}=\epsilon/d$,
Eq.\ \eqref{nC} numerically reads%
\begin{equation}
n_{C}=\frac{\epsilon V_{g}}{ed}=\frac{\epsilon_{r}V_{g}}{d}\times
5.\,\allowbreak526\,3\times10^{12}%
%TCIMACRO{\unit{cm}}%
%BeginExpansion
\operatorname{cm}%
%EndExpansion
^{-2}, \label{nC uniform}%
\end{equation}
where $V_{g}$ and $d$ are in units of V and nm, respectively.

\subsubsection{Using pdetool\label{sec pdetool CCM}}

The Dirichlet boundary conditions%
\begin{equation}
u(x,z)=%
\begin{cases}
0, & \text{at graphene boundary}\\
V_{g}, & \text{at gate boundary}%
\end{cases}
\label{Dirichlet BS for CCM}%
\end{equation}
can be straightforwardly implemented in the pdetool.\footnote{If a uniform
capacitor (without $x$ dependence) is desired, one needs to assign Neumann
boundary conditions at the left and right sides of the oxide boundaries with
vanishing surface charge density $g=0$, which forces the displacement field to
be tangential (normal) to the side (top and bottom) boundaries.} The standard
PDE solver \texttt{assempde} introduced in Sec.\ \ref{sec commands} should be
chosen. Working with units $%
%TCIMACRO{\unit{V}}%
%BeginExpansion
\operatorname{V}%
%EndExpansion
$ and $%
%TCIMACRO{\unit{nm}}%
%BeginExpansion
\operatorname{nm}%
%EndExpansion
$, the carrier density
%TCIMACRO{\TeXButton{EQref}{\eqref{nC}} }%
%BeginExpansion
\eqref{nC}
%EndExpansion
is numerically given by%
\begin{equation}
n_{C}(x)=\epsilon_{r}\left.  \frac{\partial u(x,z)}{\partial z}\right\vert
_{z=0}\times5.\,\allowbreak526\,3\times10^{12}%
%TCIMACRO{\unit{cm}}%
%BeginExpansion
\operatorname{cm}%
%EndExpansion
^{-2}. \label{nC numerical}%
\end{equation}
Note that the interpolation command\ \texttt{tri2grid} introduced in
Sec.\ \ref{sec commands} may be useful in performing the numerical derivative
$\partial u/\partial z$ at $z=0$.

\subsubsection{Remark on the gate-induced Rashba spin splitting}

At this stage we may also estimate for graphene the gate-induced Rashba spin
splitting, an intrinsic coupling between the spin and orbital degrees of
freedom of charge carriers in a two-dimensional conducting plane subject to a
perpendicular electric field \cite{Rashba1960,Bychkov1984}. In graphene, the
Rashba spin splitting has been shown by first principles to exhibit a linear
dependence on the electric field \cite{Gmitra2009,Abdelouahed2010}:
$\Delta_{R}\approx0.01\left\vert E_{e}\right\vert
%TCIMACRO{\unit{meV}}%
%BeginExpansion
\operatorname{meV}%
%EndExpansion
$, where $E_{e}$ is the electric field strength perpendicular to graphene
given in units of $%
%TCIMACRO{\unit{V}}%
%BeginExpansion
\operatorname{V}%
%EndExpansion
/%
%TCIMACRO{\unit{nm}}%
%BeginExpansion
\operatorname{nm}%
%EndExpansion
$. If the graphene carrier density stems from gating and is classically given
by $n$, the corresponding surface charge density $\left\vert \rho
_{s}\right\vert =e\left\vert n\right\vert =\epsilon\left\vert E_{e}\right\vert
$, in fact, has already revealed the displacement field on itself, allowing us
to express the Rashba spin splitting in terms of the carrier density,
\begin{equation}
\Delta_{R}=\frac{n}{\epsilon_{r}}\times1.\,\allowbreak809\,5\times10^{-6}%
%TCIMACRO{\unit{eV}}%
%BeginExpansion
\operatorname{eV}%
%EndExpansion
,\label{rashba}%
\end{equation}
where $n$ is in units of $10^{12}%
%TCIMACRO{\unit{cm}}%
%BeginExpansion
\operatorname{cm}%
%EndExpansion
^{-2}$, a typical order of the graphene carrier density.

This estimation indicates that the Rashba spin splitting induced solely by
electric gating typically lies in the order of $\mu$eV, which may hinder the
observation of those interesting physics based on the Rashba spin-orbit
coupling in graphene, such as the interfacial spin and charge currents
\cite{Yamakage2011,Tian2012}, or the spin-dependent Klein tunneling
\cite{Yamakage2009,Liu2012}. The position dependence of the Rashba coupling
across a \textit{pn} junction interface \cite{Rataj2011}, on the other hand,
can be accurately taken into account by putting the $x$-dependence of $n$ (or
even $\epsilon_{r}$) in Eq.\ \eqref{rashba}. A stronger Rashba spin splitting
in graphene is therefore less possible by gating, but may be achieved by, for
example, using a ferromagnetic substrate with an intercalated gold monolayer
\cite{Sanchez-Barriga2010}.

\subsection{Self-consistent Poisson-Dirac iteration method}

From Eq.\ \eqref{rhos in CM} to Eq.\ \eqref{nC}, the assumption that
\textquotedblleft the induced surface charge density at the dielectric surface
is the graphene carrier density\textquotedblright\ obviously have neglected a
few physical details, such as the graphene density of states that govern the
statistics of how the states in graphene should be filled by the carriers
accordingly. In addition, filling the carriers into graphene causes the change
of its Fermi level, implying a potential energy shift that should further
correspond to the electric potential times the electron charge. These are what
the classical capacitance model have neglected and what the following
Poisson-Dirac iteration method is going to compensate.

\subsubsection{Basic idea\label{sec basic idea of PDM}}

Consider a pristine graphene with Fermi level lying exactly at the charge
neutrality point, i.e., the Dirac point $E_{F}=0$. Application of the gate
voltage $V_{g}$ induces a certain amount of additional charges on graphene,
$\rho_{s}=-en$, which occupy the states in graphene according to its density
of states $D(E)=2\left\vert E\right\vert /\pi\left(  \hbar v_{F}\right)  ^{2}$
(within the Dirac model):%
\begin{equation}
n(E)=\int_{-\infty}^{E}D(E^{\prime})dE^{\prime}=\operatorname*{sgn}(E)\frac
{1}{\pi}\left(  \frac{E}{\hbar v_{F}}\right)  ^{2}, \label{n(E)}%
\end{equation}
where $v_{F}\approx10^{8}%
%TCIMACRO{\unit{cm}}%
%BeginExpansion
\operatorname{cm}%
%EndExpansion
/%
%TCIMACRO{\unit{s}}%
%BeginExpansion
\operatorname{s}%
%EndExpansion
$ is the Fermi velocity of graphene. A positive (negative) electron number
density $n$ raises (lowers) the Fermi level from $0$ to $E$. On the other
hand, the electron at the Fermi level, which is responsible for transport in
the linear response regime, gains an energy $-eV_{G}$ from the electric field,
where $-e$ is the electron charge\footnote{Throughout this paper,
$e=1.60217733\times10^{-19}%
%TCIMACRO{\unit{C}}%
%BeginExpansion
\operatorname{C}%
%EndExpansion
$ is the positive elementary charge.} and $V_{G}$ is the electric potential at
the graphene sheet obtained by solving the Poisson
%TCIMACRO{\TeXButton{EQref}{Eq.\ \eqref{poisson eq}}}%
%BeginExpansion
Eq.\ \eqref{poisson eq}%
%EndExpansion
. This potential energy $-eV_{G}$, which is equivalent to the
\textquotedblleft on-site energy\textquotedblright\ in the tight-binding
transport formulation (see, for example, \cite{Liu2012a}), will raise the
whole band structure, and thus lower the Fermi level by the same amount. We
can therefore legitimately put%
\begin{equation}
E=-(-eV_{G})=+eV_{G} \label{E=eVG}%
\end{equation}
into Eq.\ \eqref{n(E)}, leading to%
\begin{equation}
\frac{\rho_{s}}{\epsilon_{0}}=\frac{-en}{\epsilon_{0}}=-\frac{e}{\epsilon_{0}%
}\operatorname*{sgn}(V_{G})\frac{(eV_{G})^{2}}{\pi(\hbar v_{F})^{2}}.
\label{rhos in PDM}%
\end{equation}
The surface charge density at the graphene layer is now expressed in terms of
the solution $u(x,z=0)$, but is at the same time the Neumann boundary
condition that influences the numerical solution to the Poisson equation. This
formally makes the solution process iterative.

\subsubsection{Using pdetool\label{sec pdetool PDM}}

The Dirichlet boundary condition \eqref{Dirichlet BS for CCM} for the gate
boundary remains valid, while that for the graphene boundary has to be
modified to the Neumann type:%
\begin{equation}
u(x,z)=%
\begin{cases}
g, & \text{at graphene boundary}\\
V_{g}, & \text{at gate boundary}%
\end{cases}
, \label{BS for PDM}%
\end{equation}
where $g=\rho_{s}/\epsilon_{0}$ is given by Eq.\ \eqref{rhos in PDM}. Working
with units $%
%TCIMACRO{\unit{V}}%
%BeginExpansion
\operatorname{V}%
%EndExpansion
$ and $%
%TCIMACRO{\unit{nm}}%
%BeginExpansion
\operatorname{nm}%
%EndExpansion
$ together with $v_{F}=10^{8}%
%TCIMACRO{\unit{cm}}%
%BeginExpansion
\operatorname{cm}%
%EndExpansion
/%
%TCIMACRO{\unit{s}}%
%BeginExpansion
\operatorname{s}%
%EndExpansion
$, Eq.\ \eqref{rhos in PDM} becomes%
\begin{equation}
\frac{\rho_{s}}{\epsilon_{0}}=-13.\,\allowbreak295\operatorname*{sgn}%
(V_{G})\left(  \frac{V_{G}}{%
%TCIMACRO{\unit{V}}%
%BeginExpansion
\operatorname{V}%
%EndExpansion
}\right)  ^{2}\frac{%
%TCIMACRO{\unit{V}}%
%BeginExpansion
\operatorname{V}%
%EndExpansion
}{%
%TCIMACRO{\unit{nm}}%
%BeginExpansion
\operatorname{nm}%
%EndExpansion
}, \label{g in PDM}%
\end{equation}
which should be keyed as \textquotedblleft%
\texttt{-13.295*sign(u).*u.\symbol{94}2}\textquotedblright\ in the boundary
condition matrix, noting that the solution in the pdetool is by default named
$u$. The nonlinear solver \texttt{pdenonlin} introduced in
Sec.\ \ref{sec commands} has to be chosen in this case, where the solution is
involved in the boundary conditions, and the iteration will be automatically
processed by the pdetool.

Once the solution $u(x,z)$, and hence the electrostatic potential at the
graphene layer $V_{G}(x)=u(x,z=0)$, is iteratively obtained, the desired
carrier density profile $n(x)$ can then be expressed in terms of $V_{G}(x)$:%
\begin{equation}
n_{\text{PD}}(x)=7.\,\allowbreak347\,1\times10^{13}\times\operatorname*{sgn}%
[V_{G}(x)]\left[  \frac{V_{G}(x)}{%
%TCIMACRO{\unit{V}}%
%BeginExpansion
\operatorname{V}%
%EndExpansion
}\right]  ^{2}%
%TCIMACRO{\unit{cm}}%
%BeginExpansion
\operatorname{cm}%
%EndExpansion
^{-2}, \label{nPD}%
\end{equation}
which follows from Eqs.\ \eqref{n(E)} and \eqref{E=eVG}. Note that we have
added explicitly a subscript \textquotedblleft PD\textquotedblright\ in
Eq.\ \eqref{nPD} to distinguish with the classical contribution, $n_{C}$.

\subsection{Quantum capacitance model}

The relation between the induced charge density on graphene and the electric
potential energy that those charge carriers gain through the graphene density
of states is taken into account by the PDM, with a price of iteration process
paid. For single-gated graphene, there is an alternative that can take this
into account analytically: the quantum capacitance model \cite{Luryi1988},
which we briefly review here for bulk graphene following the work of
\cite{Fang2007}.\footnote{For the general case of multigated doped graphene,
see \cite{Liu2013}. The derivation is similar, and the review here is
restricted to the simple case of single-gated pristine graphene.}

\subsubsection{The model}

The single-gated graphene shown in Fig.\ \ref{fig1a} is treated by the
equivalent circuit plot as shown in Fig.\ \ref{fig1c}, where an additional
capacitor $C_{Q}$ is inserted between the voltage point $V_{G}$ and the
ground, as contrary to the CCM, Fig.\ \ref{fig1b}. As Fig.\ \ref{fig1c}
suggests $V_{g}=V_{G}+V_{\text{ox}}$, using $C_{\text{ox}}=\left\vert
\rho\right\vert /V_{\text{ox}}=en/V_{\text{ox}}$ we have%
\begin{equation}
V_{g}=V_{G}+\frac{en}{C_{\text{ox}}}\implies n=\underset{\text{classical}%
}{\underbrace{\left(  \frac{C_{\text{ox}}}{e}V_{g}\right)  }}+\underset
{\text{quantum}}{\underbrace{\left(  -\frac{C_{\text{ox}}}{e}V_{G}\right)  }%
}.\label{n from V}%
\end{equation}
Following the same physics stated in Sec.\ \ref{sec basic idea of PDM}, the
carrier density at the graphene layer, i.e., Eq.\ \eqref{rhos in PDM} divided
by $-e/\epsilon_{0}$, is expressed in terms of the electric potential thereof
as
\begin{equation}
n=\operatorname*{sgn}(eV_{G})\frac{1}{\pi}\left(  \frac{eV_{G}}{\hbar v_{F}%
}\right)  ^{2}.\label{n from DOS}%
\end{equation}
Equating \eqref{n from V} and \eqref{n from DOS}, one obtains a quadratic
equation for $V_{G}$,%
\begin{equation}
\operatorname*{sgn}(eV_{G})\frac{1}{\pi}\left(  \frac{eV_{G}}{\hbar v_{F}%
}\right)  ^{2}=\frac{C_{\text{ox}}}{e}V_{g}-\frac{C_{\text{ox}}}{e}%
V_{G}.\label{quadratic eq for VG}%
\end{equation}
Solving Eq.\ \eqref{quadratic eq for VG} for $V_{G}$ and putting back to
Eq.\ \eqref{n from V}, the graphene carrier density can be written as%
\begin{equation}
n=n_{C}+\Delta n,\label{n}%
\end{equation}
where $n_{C}$ given by Eq.\ \eqref{nC} is the classical contribution, and%
\begin{equation}
\Delta n=\operatorname*{sgn}(n_{C})n_{Q}\left(  1-\sqrt{1+2\frac{\left\vert
n_{C}\right\vert }{n_{Q}}}\right)  ,\label{Delta n}%
\end{equation}
with{} definition%
\begin{equation}
n_{Q}=\frac{\pi}{2}\left(  \frac{C_{\text{ox}}\hbar v_{F}}{e^{2}}\right)
^{2},\label{nQ}%
\end{equation}
corresponds to the quantum correction.\footnote{Note that
Eqs.\ \eqref{n}--\eqref{nQ} (with $n_{C}>0$) were first derived in
\cite{Fang2007} and reviewed in \cite{DasSarma2011}, but a factor of $2$ in
the square root of the formula (1.15) in \cite{DasSarma2011}, corresponding to
Eq.\ \eqref{Delta n} here, is missing.}

Furthermore, by comparing Eq.\ \eqref{n} with Eq.\ \eqref{n from V}, one can
also write down the solution for the electric potential on graphene:%
\begin{equation}
V_{G}=-\frac{e\Delta n}{C_{\text{ox}}}=-\frac{\operatorname*{sgn}(n_{C}%
)n_{Q}\left(  1-\sqrt{1+2\dfrac{\left\vert n_{C}\right\vert }{n_{Q}}}\right)
}{C_{\text{ox}}/e},\label{VG by QCM}%
\end{equation}
which has a reasonable form of charge divided by capacitance. The
\textquotedblleft charge\textquotedblright\ in Eq.\ \eqref{VG by QCM} contains
only the quantum correction as expected, since the classical solution, which
regards graphene as a grounded conductor, does not contribute to the potential
$V_{G}$.\begin{figure}[b]
\centering\includegraphics[width=\columnwidth]{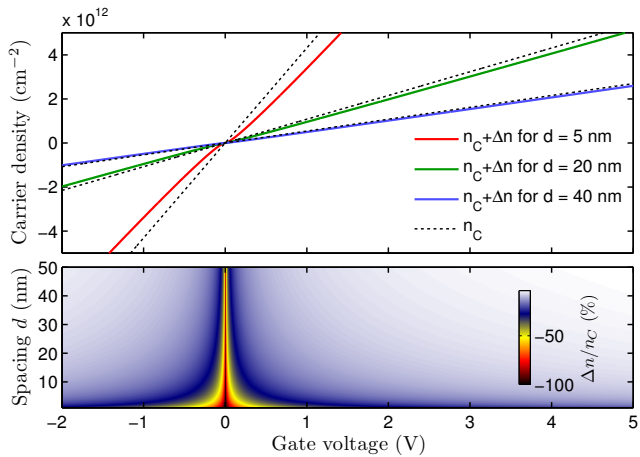}\caption{Upper panel:
The carrier density as a function of gate voltage $V_{g}$, with and without
the quantum correction $\Delta n$, considering oxide thickness
$d=5,20,40\operatorname{nm}$ with $\epsilon_{r}=3.9$ of the dielectric
SiO$_{2}$ assumed. Lower panel: Two-dimensional color plot of $\Delta n/n_{C}$
as a function of $V_{g}$ and $d$.}%
\label{fig2}%
\end{figure}

\subsubsection{Quantum correction for parallel-plate
capacitors\label{sec quantum correction}}

For parallel-plate capacitors with uniform dielectrics, the capacitance is
well known to be $C_{\text{ox}}=\epsilon/d$, such that Eq.\ \eqref{nQ} and
hence the quantum correction \eqref{Delta n} are solely determined by the
classical contribution $n_{C}=(C_{\text{ox}}/e)V_{g}$. In this case we can
further express Eq.\ \eqref{nQ} as $n_{Q}=\left(  \epsilon_{r}/d\right)
^{2}\times2.\,\allowbreak078\,4\times10^{11}%
%TCIMACRO{\unit{cm}}%
%BeginExpansion
\operatorname{cm}%
%EndExpansion
^{-2}$, where $d$ is in units of nm and $v_{F}=10^{8}%
%TCIMACRO{\unit{cm}}%
%BeginExpansion
\operatorname{cm}%
%EndExpansion
/%
%TCIMACRO{\unit{s}}%
%BeginExpansion
\operatorname{s}%
%EndExpansion
$ is again adopted. Together with Eq.\ \eqref{nC uniform}, the quantum
correction given by Eq.\ \eqref{Delta n} can be written as
\begin{equation}%
\begin{split}
\Delta n  &  =\operatorname*{sgn}(V_{g})\left(  \frac{\epsilon_{r}}{d}\right)
^{2}\left(  1-\sqrt{1+53.\,\allowbreak178\frac{\left\vert V_{g}\right\vert
d}{\epsilon_{r}}}\right) \\
&  \times2.\,\allowbreak078\,4\times10^{11}%
%TCIMACRO{\unit{cm}}%
%BeginExpansion
\operatorname{cm}%
%EndExpansion
^{-2}%
\end{split}
, \label{Delta n uniform}%
\end{equation}
where $V_{g}$ is in units of V and $\operatorname*{sgn}(n_{C}%
)=\operatorname*{sgn}(C_{\text{ox}}V_{g}/e)=\operatorname*{sgn}(V_{g})$ has
been substituted. We will soon see that this correction derived within the
analytical QCM for an infinitely extending parallel-plate capacitor will
exactly correspond to that by the numerical PDM.

To give an overview of how much change the quantum correction $\Delta n$
causes as compared to the classical $n_{C}$, we plot in the upper panel of
Fig.\ \ref{fig2} the carrier density, with and without $\Delta n$, as a
function of $V_{g}$, considering oxide thickness $d=5,20,40%
%TCIMACRO{\unit{nm}}%
%BeginExpansion
\operatorname{nm}%
%EndExpansion
$ with $\epsilon_{r}=3.9$ of the assumed dielectric SiO$_{2}$. Apparently,
only when $d$ is extremely thin can one see a clear difference due to $\Delta
n$ ($d\lesssim20%
%TCIMACRO{\unit{nm}}%
%BeginExpansion
\operatorname{nm}%
%EndExpansion
$). With nonzero $V_{g}$ and large $d$, one can further approximate
Eq.\ \eqref{Delta n uniform} as%
\begin{equation}
\Delta n\approx-\operatorname*{sgn}(V_{g})\left(  \frac{\epsilon_{r}}%
{d}\right)  ^{3/2}\sqrt{\left\vert V_{g}\right\vert }\times1.\,\allowbreak
515\,6\times10^{12}%
%TCIMACRO{\unit{cm}}%
%BeginExpansion
\operatorname{cm}%
%EndExpansion
^{-2},\label{Delta n approx}%
\end{equation}
which shows a rapid decay of $\Delta n$ with $d$ to the power of $3/2$.

In the opposite limit of vanishing $V_{g}$ and thin $d$, however, the
magnitude of $\Delta n$ may become comparable with $n_{C}$. In the lower panel
of Fig.\ \ref{fig2}, the ratio $\Delta n/n_{C}$ is plotted as a function of
$V_{g}$ and $d$. As expected, in the region away from $V_{g}=0$ and
$d\lesssim20%
%TCIMACRO{\unit{nm}}%
%BeginExpansion
\operatorname{nm}%
%EndExpansion
$, the ratio is close to zero, meaning a minor role played by the quantum
correction. Contrarily, the ratio grows significantly when approaching to the
$V_{g}$ and $d$ axes,\footnote{The ratio further diverges to $\Delta
n/n_{C}\rightarrow-100\%$ at $V_{g}=0$, but at this axis both $n_{C}$ and
$\Delta n$ vanish, and $\Delta n/n_{C}$ is strictly speaking undefined.}
implying an important role played by the quantum correction.

\subsubsection{Remark on quantum capacitance}

Note that the appearance of $C_{Q}$ stems from the finite density of states
provided by the conducting layer for the electrons to occupy following the
quantum nature of the Pauli exclusion principle, and hence the name quantum
capacitance \cite{Luryi1988}, which is not restricted to the material
graphene. The expression of $C_{Q}$ for graphene \cite{Fang2007}, however, is
not important for the present discussion. Instead, $C_{Q}$ leads to a quantum
correction to the gate-induced carrier density $\Delta n$, which is the main
focus here.

In addition, recent experimental progress on the measurement of graphene
quantum capacitance \cite{Xia2009,Droscher2010,Ponomarenko2010} suggests that
the electron-hole puddles \cite{Martin2008} induced by charged impurities may
influence $C_{Q}$ at energies close to the charge neutrality point. The
corresponding carrier density fluctuation $\delta n$, which can be considered
to develop a microscopic model to account for the smoothing of the graphene
quantum capacitance at the charge neutrality point \cite{Xu2011}, is beyond
the scope of the present discussion.

\subsection{Analytics vs numerics\label{sec analytics vs numerics}}

The two analytical capacitance models and the numerical scheme of the
Poisson-Dirac iteration method are compared in the following, considering a
single-gated graphene with individually infinite and finite size of the
gate.\begin{figure}[b]
\centering\includegraphics[width=\columnwidth]{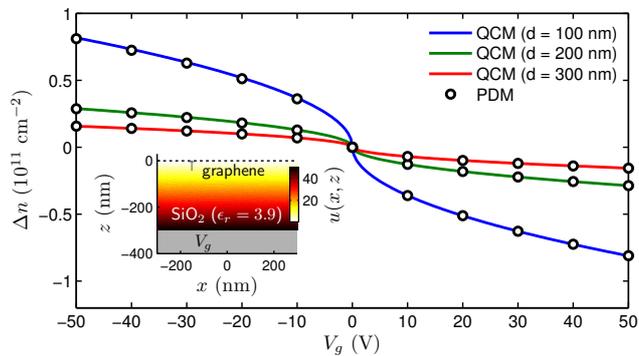}\caption{The quantum
correction to the gate-induced carrier density on graphene in the case of
parallel-plate capacitor, calculated by the analytical quantum capacitance
model and the numerical Poisson-Dirac iteration method. Inset: Schematic of
the capacitor with $d=300\operatorname{nm}$ with the color shading
representing the iterated electric potential solution $u(x,z)$ obtained by the
PDM at $V_{g}=50\operatorname{V}$.}%
\label{fig3}%
\end{figure}

We start with uniform parallel-plate capacitors as those considered in
Sec.\ \ref{sec quantum correction} with different spacings $d=100,200,300$ nm.
Schematic of the capacitor with $d=300%
%TCIMACRO{\unit{nm}}%
%BeginExpansion
\operatorname{nm}%
%EndExpansion
$ is sketched in the inset of Fig.\ \ref{fig3}, where the electric potential
$u(x,z)$ within the oxide layer is obtained by the PDM at $V_{g}=50%
%TCIMACRO{\unit{V}}%
%BeginExpansion
\operatorname{V}%
%EndExpansion
$.\footnote{Note that the spatial profile of the electric potential $u(x,z)$,
with the quantum correction on graphene taken into account, does not look too
much different compared to the classical solution $u_{0}(x,z)$, where the
graphene layer is assumed to be grounded. The difference of them at $z=0$,
however, is crucial since the latter is always zero, i.e., $u_{0}(x,z=0)=0$.}
The classical contribution $n_{C}$ is first computed following
Sec.\ \ref{sec pdetool CCM} [which gives the same result with
Eq.\ \eqref{nC uniform}], and the quantum correction $\Delta n$ is computed in
two ways. For the analytical QCM, Eq.\ \eqref{Delta n uniform} is used to
compute $\Delta n$. For the numerical PDM, the full carrier density
$n_{\text{PD}}$ is computed following Sec.\ \ref{sec pdetool PDM}, and the
correction is given by the difference $n_{\text{PD}}-n_{C}$. As shown in
Fig.\ \ref{fig3}, the correspondence between the two approaches is
exact.\begin{figure}[t]
\centering\includegraphics[width=\columnwidth]{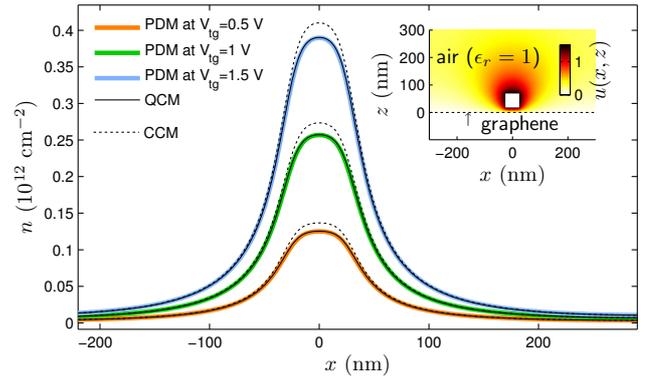}\caption{Comparison
between the position-dependent carrier densities calculated by the PDM,
$n_{\text{PD}}(x)$, the QCM, $n_{\text{QC}}(x)$, as well as CCM, $n_{C}(x)$,
at various topgate voltages. Inset: Schematic of the graphene sheet in the
presence of the topgate, suspended $20$ nm above, with cross section $50$ nm
$\times50$ nm. The spatial distribution of the electric potential $u(x,z)$ is
obtained by the iterative PDM at $V_{\text{tg}}=1.5\operatorname{V}$.}%
\label{fig4}%
\end{figure}

Next we consider a finite-size suspended topgate (such as those fabricated in
\cite{Liu2008c,Gorbachev2008}) with various voltages $V_{\text{tg}%
}=0.5,1.0,1.5%
%TCIMACRO{\unit{V}}%
%BeginExpansion
\operatorname{V}%
%EndExpansion
$ and an extremely narrow spacing $d=20$ nm; see the inset of Fig.\ \ref{fig4}%
. The calculations are similar to those for the infinite case described above.
The only difference is the approximating form of the dielectric capacitance,%
\begin{equation}
C_{\text{ox}}(x)\equiv\frac{e\cdot n_{C}(x)}{V_{\text{tg}}}, \label{Cox(x)}%
\end{equation}
from which $n_{Q}$ given by Eq.\ \eqref{nQ} and hence the quantum correction,
Eq.\ \eqref{Delta n} from the analytical QCM, is obtained. In Fig.\ \ref{fig4}%
, we compare the carrier densities calculated by the PDM, $n_{\text{PD}}$, the
QCM, $n_{\text{QC}}$, as well as the CCM, $n_{C}$. Good agreement between PDM
and QCM is again clearly seen, while the deviation of the CCM from them is
observable due to the rather thin spacing $d=20%
%TCIMACRO{\unit{nm}}%
%BeginExpansion
\operatorname{nm}%
%EndExpansion
$ between the gate and the graphene sheet.

From the above testing calculations (Figs.\ \ref{fig3} and \ref{fig4}), we may
conclude that the QCM is equivalent to the PDM in both cases of infinite
(uniform) and finite (nonuniform) gate-graphene capacitors. In particular, the
correspondence between the two approaches in the former case is exact, while
the discrepancy in the latter is merely negligible, suggesting that
Eq.\ \eqref{Cox(x)} is a good approximation for calculating the spatially
varying oxide capacitance $C_{\text{ox}}(x)$ that further determines the
quantum correction $\Delta n$ given by Eq.\ \eqref{Delta n} through the
definition \eqref{nQ} within the QCM. The classical solution $n_{C}(x)$ for
the nonuniform case [following Sec.\ \ref{sec pdetool CCM}], therefore, serves
as the preliminary solution step for the exactly solvable QCM, circumventing
the self-consistent iteration during the solution process that is needed in
the PDM.

\subsection{Beyond single-gated pristine graphene\label{sec beyond}}

The above discussion considers only single-gated graphene in the absence of
chemical doping. For double-gated graphene with topgate and backgate at two
sides of the graphene sheet, the two gates can be regarded as independent, and
their contributions to the carrier density modulation can be treated
separately and superposed to yield the total carrier density. When multiple
gates are acting on the graphene sheet from the same side, however, such as
using an embedded local gate in addition to a global backgate to create
\textit{pnp} junctions with independent control of the globally and locally
gated regions \cite{Nam2011} (see Sec.\ \ref{sec embedded}), or patterned
topgates that may generate a graphene superlattice (see
Sec.\ \ref{sec superlattice}), these gates should be simultaneously treated.

In fact, the CCM (Sec.\ \ref{sec pdetool CCM}) as well as the PDM
(Sec.\ \ref{sec pdetool PDM}) are not restricted to the case of single-gated
graphene. These two approaches work for any kind of gating geometry, provided
that the boundary conditions [Eq.\ \eqref{Dirichlet BS for CCM} for CCM and
Eq.\ \eqref{BS for PDM} for PDM] at the graphene layer are properly assigned.
The applicability of the QCM of the presently reviewed version, however,
depends then on the gating geometry. When multiple gates are acting on
graphene from the same side but connected to each other to share the same gate
voltage (as the case of Sec.\ \ref{sec superlattice}), there is effectively
only one gate. In this case, Eq.\ \eqref{Cox(x)} is still a good approximation
to account for the oxide capacitance, and the QCM can be directly applied. On
the other hand, if the multiple gates can be separately controlled (as the
case of Sec.\ \ref{sec embedded}), Eq.\ \eqref{Cox(x)} becomes insufficient
due to the need of multiple self-partial capacitances, and the QCM (of the
presently reviewed version) cannot be applied. Generalization of the model to
take into account composite gating geometry has been recently achieved
\cite{Liu2013}, but is beyond the scope of the present review.

For the general case of multigated doped graphene sheets, the boundary
condition for the PDM, Eq.\ \eqref{rhos in PDM}, as well as the analytical
expressions within the QCM, Eqs.\ \eqref{n}--\eqref{VG by QCM}, can be derived
similarly to the above reviewed theories. The interested readers are referred
to the recent work of \cite{Liu2013}.

\section{Applications\label{sec 4}}

A successful simulation for electronic transport in bulk graphene relies on
not only sophisticated computation techniques but also a realistic
\textquotedblleft potential\ profile\textquotedblright\ $V$ that is
experimentally relevant \cite{Liu2012a}. The term \textquotedblleft
potential\textquotedblright\ refers to the local potential energy added to the
system Hamiltonian when modeling for graphene electronic transport. Thus the
potential profile simply means the local energy band offset of the graphene
sheet subject to a spatially varying carrier density due to electrical gating.
This section is devoted to the application of the carrier density calculation:
the corresponding potential profile, or the local energy band offset, which is
a simple computational task but nevertheless important for graphene electronic
transport calculations. A few concrete examples will be illustrated, after a
short review of the potential profile is given.

\subsection{Potential profile (local energy band offset)}

In Sec.\ \ref{sec 3} we have introduced how to compute with a satisfactory
accuracy the graphene carrier density $n$, which is related to the quasi-Fermi
level through Eq.\ \eqref{n(E)} as $E_{F}=\operatorname*{sgn}(n)\hbar
v_{F}\sqrt{\pi|n|}$. The energy band offset then reads $V=E_{F}^{0}-E_{F}$,
where $E_{F}^{0}$ is the global Fermi level. Choosing $E_{F}^{0}=0$ (as is
usually the case and will be adopted in the rest of the calculations), the
space-resolved band offset reads \cite{Liu2012a}%
\begin{equation}%
\begin{split}
V(x)  &  =-\operatorname*{sgn}[n(x)]\hbar v_{F}\sqrt{\pi|n(x)|}\\
&  =-11.\,\allowbreak667\times\operatorname*{sgn}[n(x)]\sqrt{\frac{\left\vert
n(x)\right\vert }{10^{10}%
%TCIMACRO{\unit{cm}}%
%BeginExpansion
\operatorname{cm}%
%EndExpansion
^{-2}}}%
%TCIMACRO{\unit{meV}}%
%BeginExpansion
\operatorname{meV}%
%EndExpansion
\end{split}
, \label{V(x)}%
\end{equation}
which is termed on-site energy in the tight-binding formulation for transport calculations.

Equation \eqref{V(x)} interprets the carrier density profile $n(x)$ in terms
of the potential profile $V(x)$, and is valid for $n(x)$ computed by either
CCM, QCM, or PDM. It should be remarked, however, that the Poisson-Dirac
iterative solution to the electric potential at the graphene layer $V_{G}$
times $-e$ readily gives the desired energy band offset, and Eq.\ \eqref{V(x)}
is not needed within this approach. Likewise in the QCM, $V_{G}$ is given by
Eq.\ \eqref{VG by QCM} and its product with $-e$ also gives the desired
$V(x)$. Thus within QCM and PDM, one does not need to bother with
Eq.\ \eqref{V(x)} for obtaining the potential profile. Contrarily, the CCM
always treat the graphene sheet as a grounded conductor and therefore needs
the interpretation \eqref{V(x)}. In other cases where PDM is partly used but
the total carrier density is separately computed (such as
Sec.\ \ref{sec superlattice}), one needs Eq.\ \eqref{V(x)} as well.

\subsection{Graphene \textit{pnp} junctions\label{sec embedded}}

\begin{figure}[t]
\subfigure[]{
\includegraphics[width=\columnwidth]{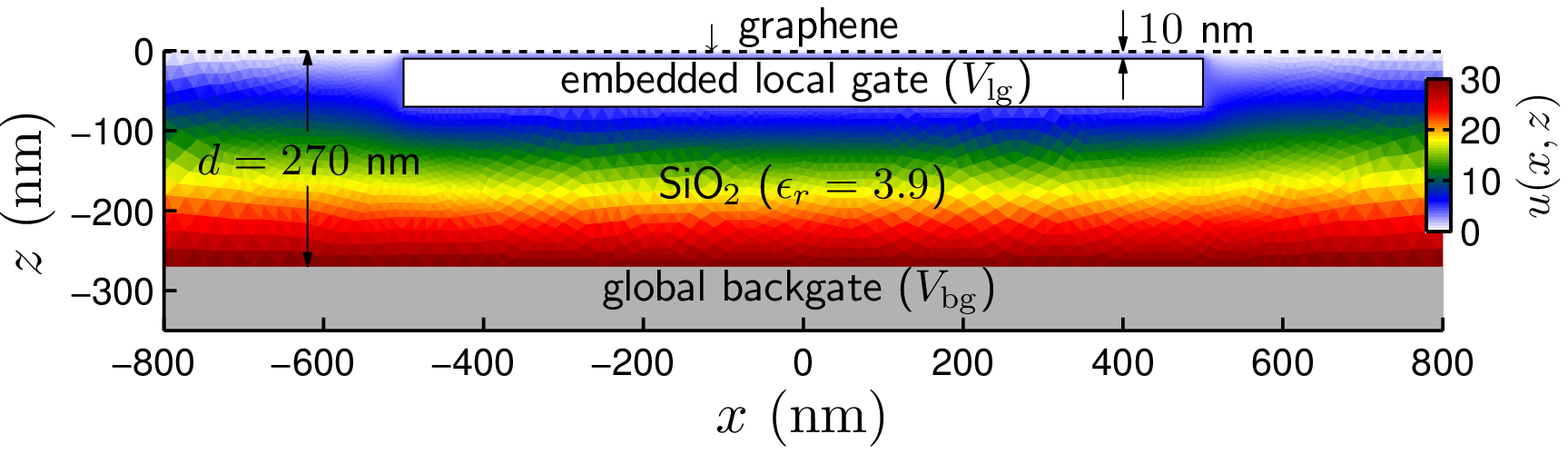}
\label{fig5a}}
\par
\subfigure[]{
\includegraphics[height=3.4cm]{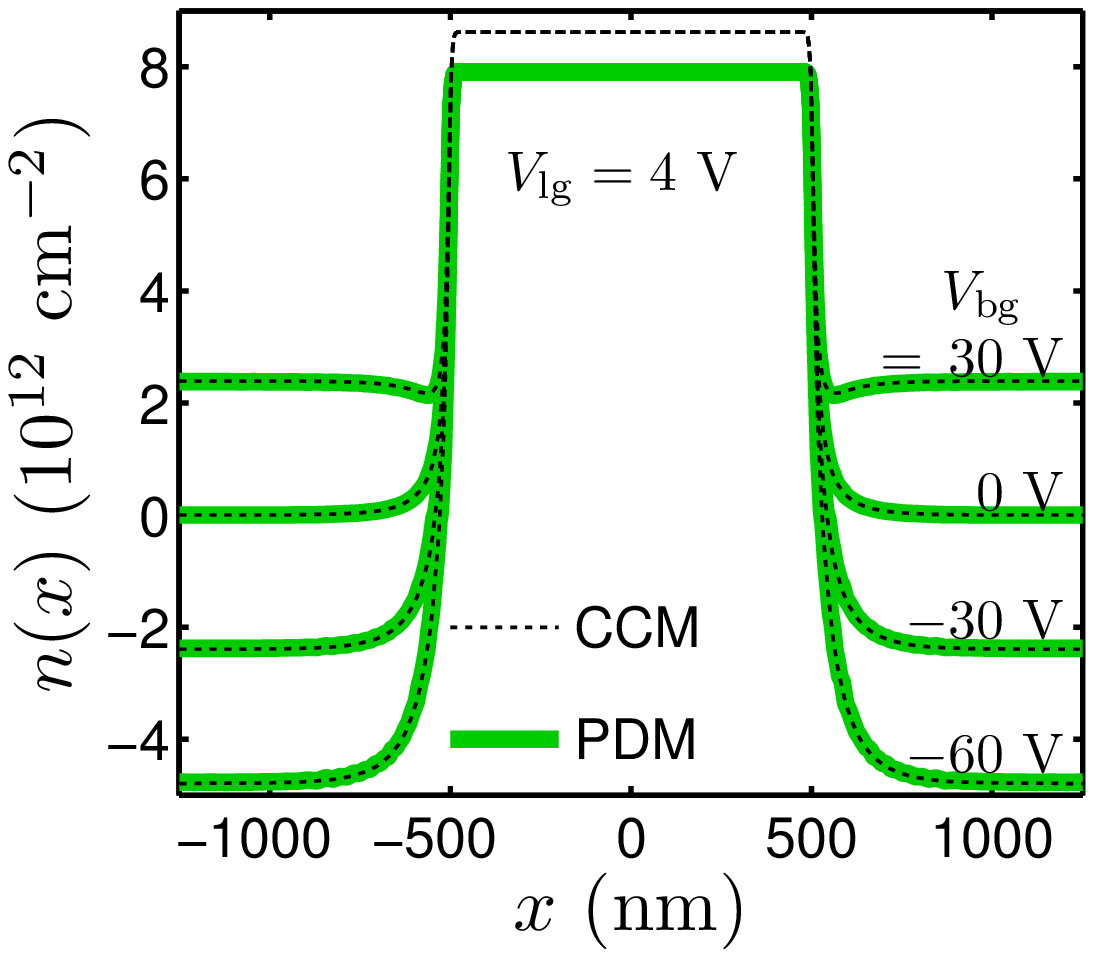}
\label{fig5b}} \hfill\subfigure[]{
\includegraphics[height=3.4cm]{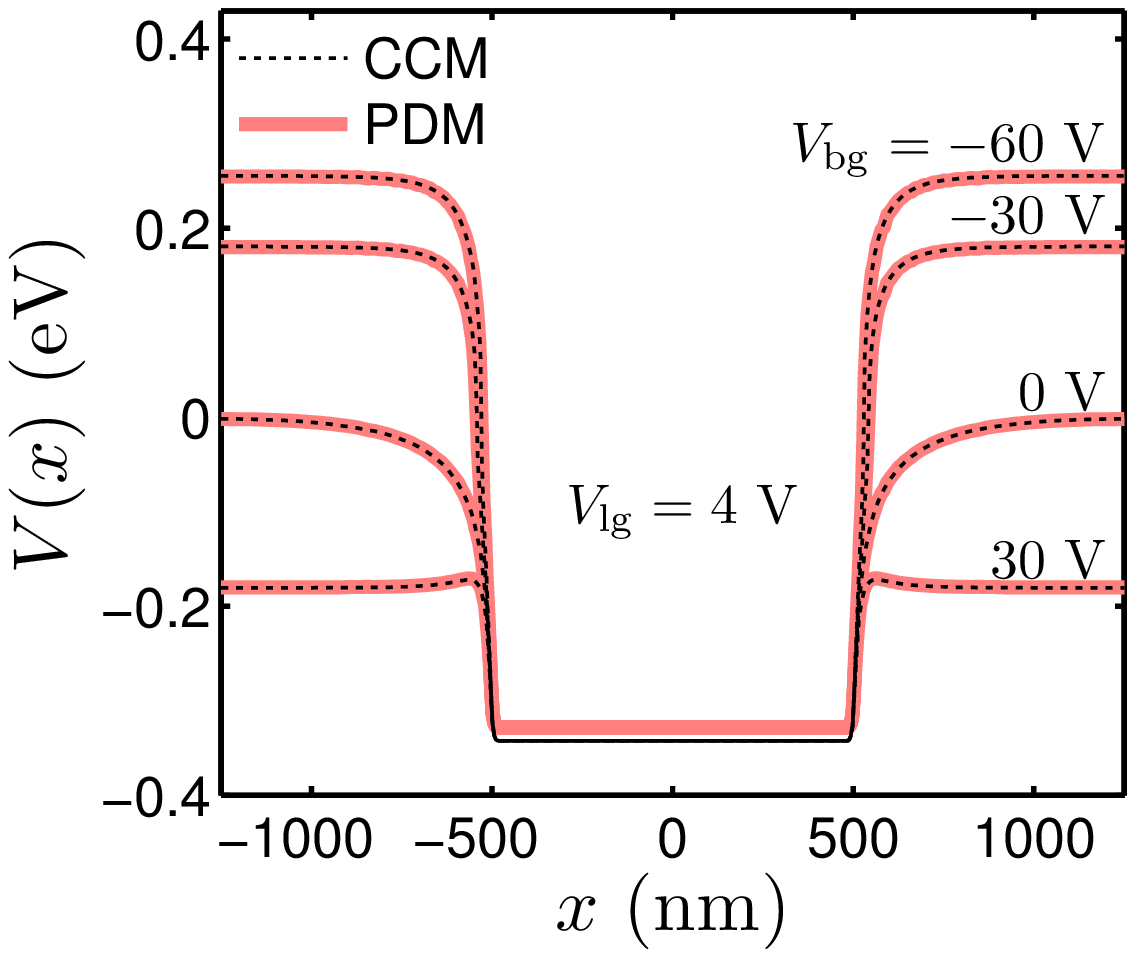}
\label{fig5c}}\caption{A graphene \textit{pnp} junction using a global
backgate with various voltages $V_{\text{bg}}$ and an embedded local gate
fixed at $V_{\text{lg}}=4$ V. (a) The iterative solution $u(x,z)$ in units of
V to the electrostatic potential within the oxide, subject to $V_{\text{bg}%
}=30$ V. The induced graphene carrier densities (b) and the corresponding
potential profiles (c) based on the CCM and PDM show that the locally gated
region $\left\vert x\right\vert \leq500\operatorname{nm}$ is not affected by
the backgate but only the local gate.}%
\label{fig5}%
\end{figure}

We begin the illustrative examples with a graphene \textit{pnp} junction,
using a global backgate and an embedded local gate. The gating geometry is
sketched in Fig.\ \ref{fig5a}, similar to those experimentally fabricated in
\cite{Nam2011}. In this case both of the global and local gates influence the
graphene carrier density from the same side, and therefore have to be treated
at the same time. As remarked previously in Sec.\ \ref{sec beyond}, the
single-gate version of the QCM does not apply here,\footnote{The multigate
version of the QCM \cite{Liu2013}, which requires to compute the self-partial
capacitances $C_{\text{lg}}$ and $C_{\text{bg}}$ due to respectively the local
gate and the backgate, can be shown to yield results well agreeing with the
PDM.} but nevertheless can be used to estimate the quantum correction due to
the embedded local gate at the region above it. Here we will mainly compare
the results from the CCM and those from the PDM, fixing the local gate voltage
at $V_{\text{lg}}=4%
%TCIMACRO{\unit{V}}%
%BeginExpansion
\operatorname{V}%
%EndExpansion
$ while varying the backgate voltage with $V_{\text{bg}}=-60,-30,0,30%
%TCIMACRO{\unit{V}}%
%BeginExpansion
\operatorname{V}%
%EndExpansion
$.

The computed carrier densities are shown in Fig.\ \ref{fig5b}, where the
Poisson-Dirac solution agrees well with the simulation presented in
\cite{Nam2011}. Since the local gate is embedded only $10%
%TCIMACRO{\unit{nm}}%
%BeginExpansion
\operatorname{nm}%
%EndExpansion
$ below the graphene sheet, the quantum correction excluded in the CCM becomes
pronounced within the locally gated region, and can be estimated by
Eq.\ \eqref{Delta n uniform} of the QCM. At the center of the locally gated
region, the electric field generated by the backgate is almost completely
screened, and the classical contribution to the carrier density can be
estimated by $n_{C}(x=0)=\epsilon V_{\text{lg}}/ed=8.\,\allowbreak
621\,1\times10^{12}%
%TCIMACRO{\unit{cm}}%
%BeginExpansion
\operatorname{cm}%
%EndExpansion
^{-2}$, leading to $n_{Q}=3.\,\allowbreak161\,2\times10^{10}%
%TCIMACRO{\unit{cm}}%
%BeginExpansion
\operatorname{cm}%
%EndExpansion
^{-2}$ and hence $\Delta n=-7.\,\allowbreak073\,5\times10^{11}%
%TCIMACRO{\unit{cm}}%
%BeginExpansion
\operatorname{cm}%
%EndExpansion
^{-2}$, which is pretty close to $n_{\text{PD}}(x=0)-n_{C}(x=0)=-7.0746\times
10^{11}%
%TCIMACRO{\unit{cm}}%
%BeginExpansion
\operatorname{cm}%
%EndExpansion
^{-2}$ from the data of Fig.\ \ref{fig5b} for all $V_{\text{bg}}$.

The carrier density profiles $n(x)$ of Fig.\ \ref{fig5b} are translated into
$V(x)$ via Eq.\ \eqref{V(x)}, as shown in Fig.\ \ref{fig5c}. The positive
$V_{\text{lg}}$ charges the locally gated graphene with a positive number of
electrons, forming an \textit{n}-type region with positive quasi-Fermi level
$E_{F}(x)>0$ that is equivalent to applying a negative energy band offset
$V(x)<0$. Outside the locally gated region, the carrier type of graphene is
controlled by the backgate with a similar principle. The most interesting
feature here is that the locally gated region can be controlled independently
due to the screening of the embedded local gate, as is evident in both
Figs.\ \ref{fig5b} and \ref{fig5c}. This independent control leads to the four
quadrants of the conductance map $G(V_{\text{lg}},V_{\text{bg}})$ with two
boundaries perpendicular to each other \cite{Nam2011}, as contrary to those
observed in top-gated devices
\cite{Huard2007,Williams2007,Ozyilmaz2007,Liu2008c,Gorbachev2008,Stander2009,Young2009}%
.\begin{figure}[t]
\subfigure[]{
\includegraphics[width=\columnwidth]{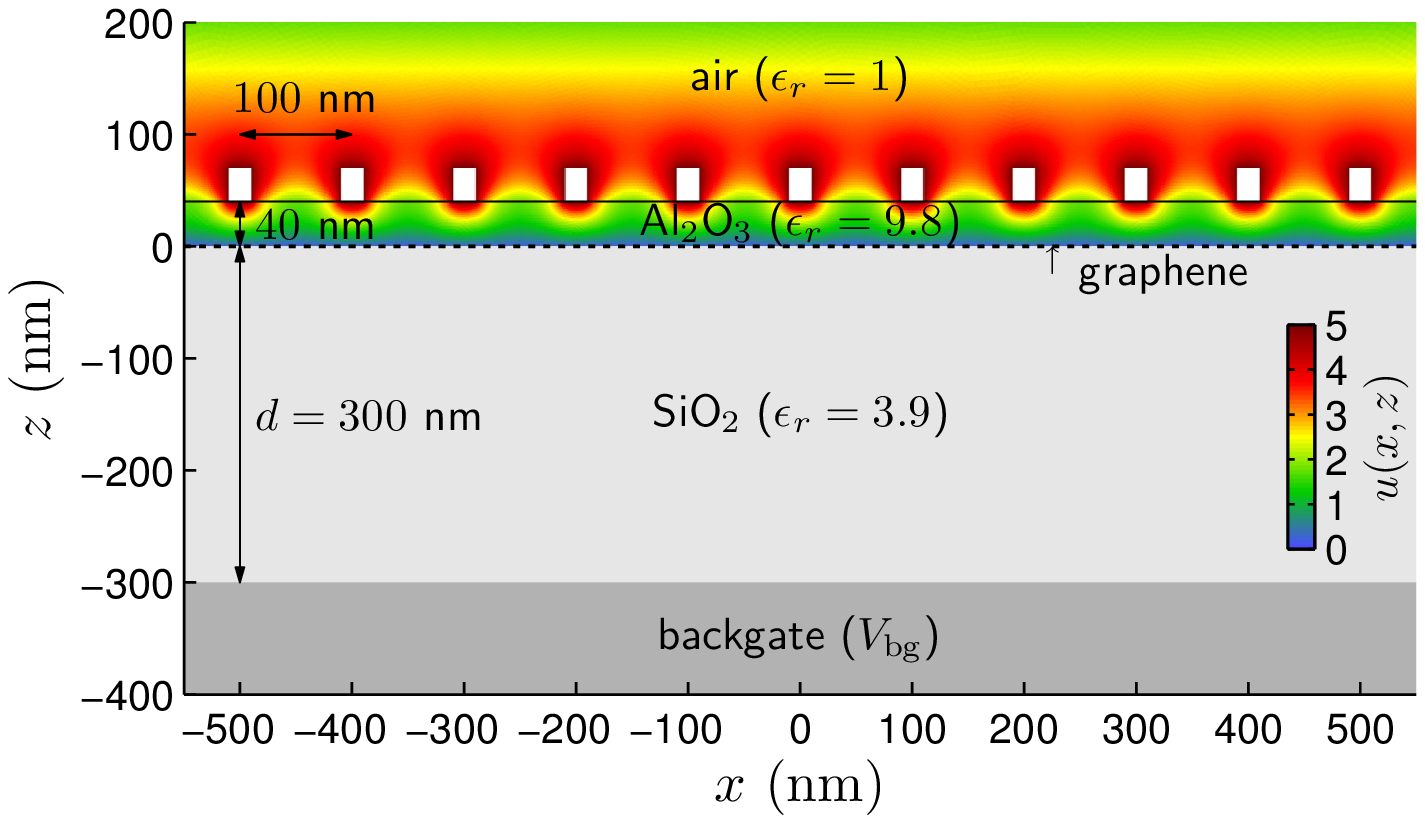}
\label{fig6a}} \subfigure[]{
\includegraphics[width=\columnwidth]{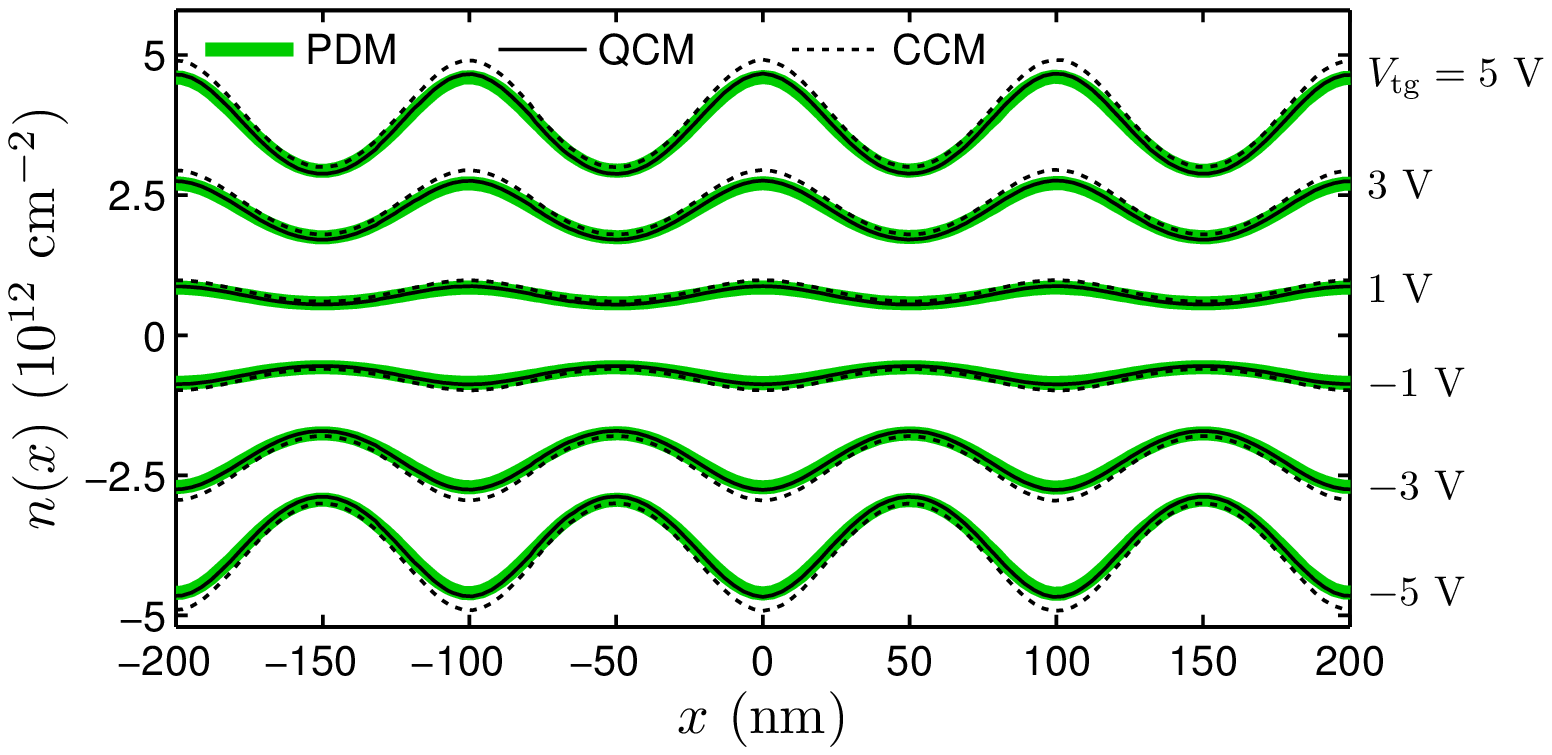}
\label{fig6b}} \subfigure[]{
\includegraphics[width=\columnwidth]{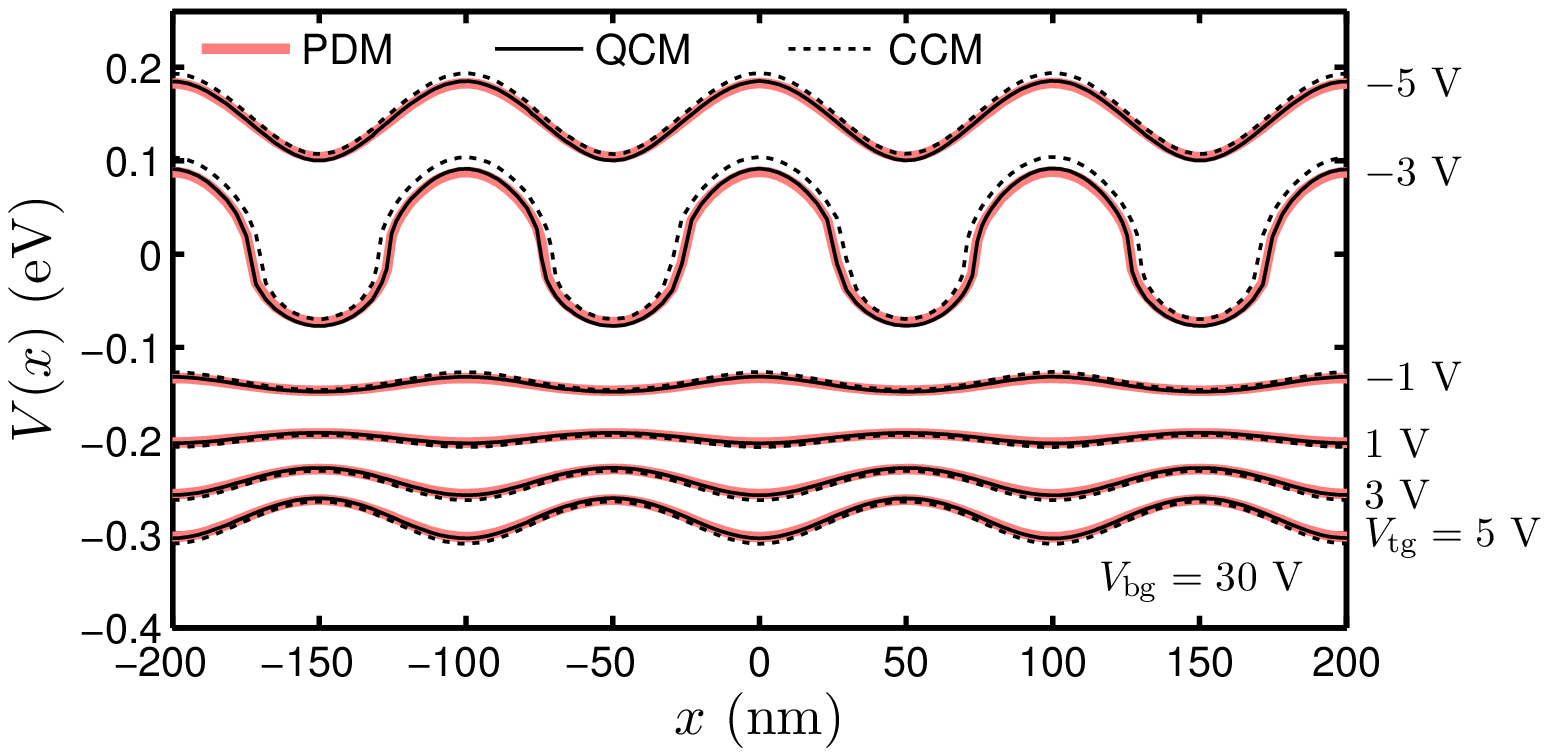}
\label{fig6c}} \caption{Formation of a graphene superlattice using a series of
patterned topgates and a global backgate, which are treated separately. (a)
Iterative solution $u(x,z)$ with $V_{\text{tg}}=5$ V, considering only the
topgates. (b) The corresponding carrier densities based on all of the three
approaches. (c) The profiles of the energy band offset from the total carrier
density composed of the patterned topgates and the uniform backgate
contributions.}%
\label{fig6}%
\end{figure}

\subsection{Graphene superlattices\label{sec superlattice}}

Next we turn to the possibility of generating a graphene superlattice by
fabricating a series of patterned topgates. As sketched in Fig.\ \ref{fig6a},
the PDE problem is defined within the region above the graphene sheet in order
to solve the electric potential due to the topgates with various voltages
$V_{\text{tg}}=-5,-3,\cdots,5%
%TCIMACRO{\unit{V}}%
%BeginExpansion
\operatorname{V}%
%EndExpansion
$. The contribution from the backgate is assumed to be uniform and can be
treated independently. The strategy here is to compute first the carrier
density due to topgates, and then include the backgate contribution to yield
the total carrier density that finally gives the energy band offset profile
from Eq.\ \eqref{V(x)}. In this case the approximation \eqref{Cox(x)} for the
QCM is rather acceptable, and we will compare the results from all of the
three approaches.

The computed carrier densities by PDM, QCM, and CCM are shown in
Fig.\ \ref{fig6b}. The curves of $n_{\text{PD}}$ and $n_{\text{QC}}$ almost
coincide with each other. The relatively thick $40%
%TCIMACRO{\unit{nm}}%
%BeginExpansion
\operatorname{nm}%
%EndExpansion
$ of Al$_{2}$O$_{3}$ suppresses the quantum correction to a reasonably small
amount, such that here the CCM is not a bad approximation, either. The
discrepancy between $n_{C}$ and $n_{\text{PD}}$ (or $n_{\text{QC}}$) is less
pronounced at the regions between each adjacent pair of topgates since the
quantum correction $\left\vert \Delta n\right\vert $ roughly decreases with
the $3/2$-th power of the distance to the gate, as mentioned in Eq.\ \eqref{Delta n approx}.

The carrier density modulation follows the patterned topgates with a
periodicity of $100%
%TCIMACRO{\unit{nm}}%
%BeginExpansion
\operatorname{nm}%
%EndExpansion
$, giving rise to a periodic potential profile $V(x)$ as shown in
Fig.\ \ref{fig6c}, where a backgate contribution with $V_{\text{bg}}=30%
%TCIMACRO{\unit{V}}%
%BeginExpansion
\operatorname{V}%
%EndExpansion
$ is taken into account. Since $V(x)$ is related to $n(x)$ through a
square-root relation, Eq.\ \eqref{V(x)}, the shape of $V(x)$ can be a bit
different from that of $n(x)$, which is similar to a sine-like wave,
especially for those $V(x)$ that alternate between positive and negative
values. Note that the backgate voltage chosen in Fig.\ \ref{fig6c} results in
a rather symmetric $V_{\text{tg}}=-3%
%TCIMACRO{\unit{V}}%
%BeginExpansion
\operatorname{V}%
%EndExpansion
$ curve since the corresponding carrier density $n(x)$ alternates
symmetrically between positive and negative at this combination of gate
voltages. In general, the alternation of $n(x)$ is not necessarily symmetric
(about the charge neutrality point $n=0$), and the resulting $V(x)$ profile
can be of peculiar shapes.

\subsection{Linear potential}

\begin{figure}[t]
\subfigure[]{
\includegraphics[width=\columnwidth]{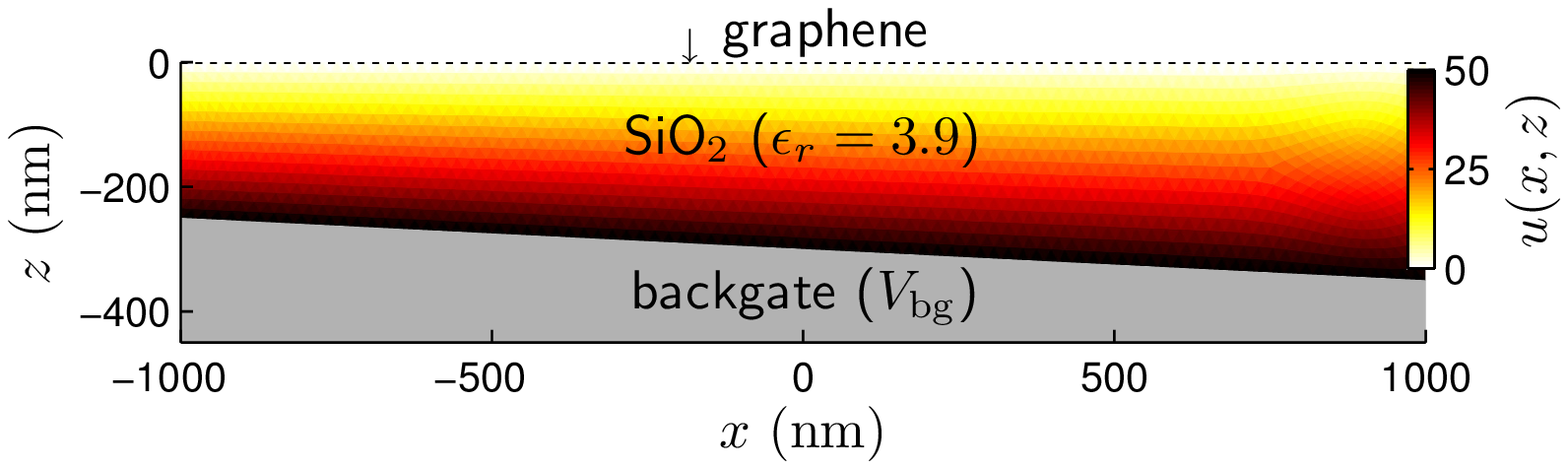}
\label{fig7a}}
\par
\subfigure[]{
\includegraphics[height=3.3cm]{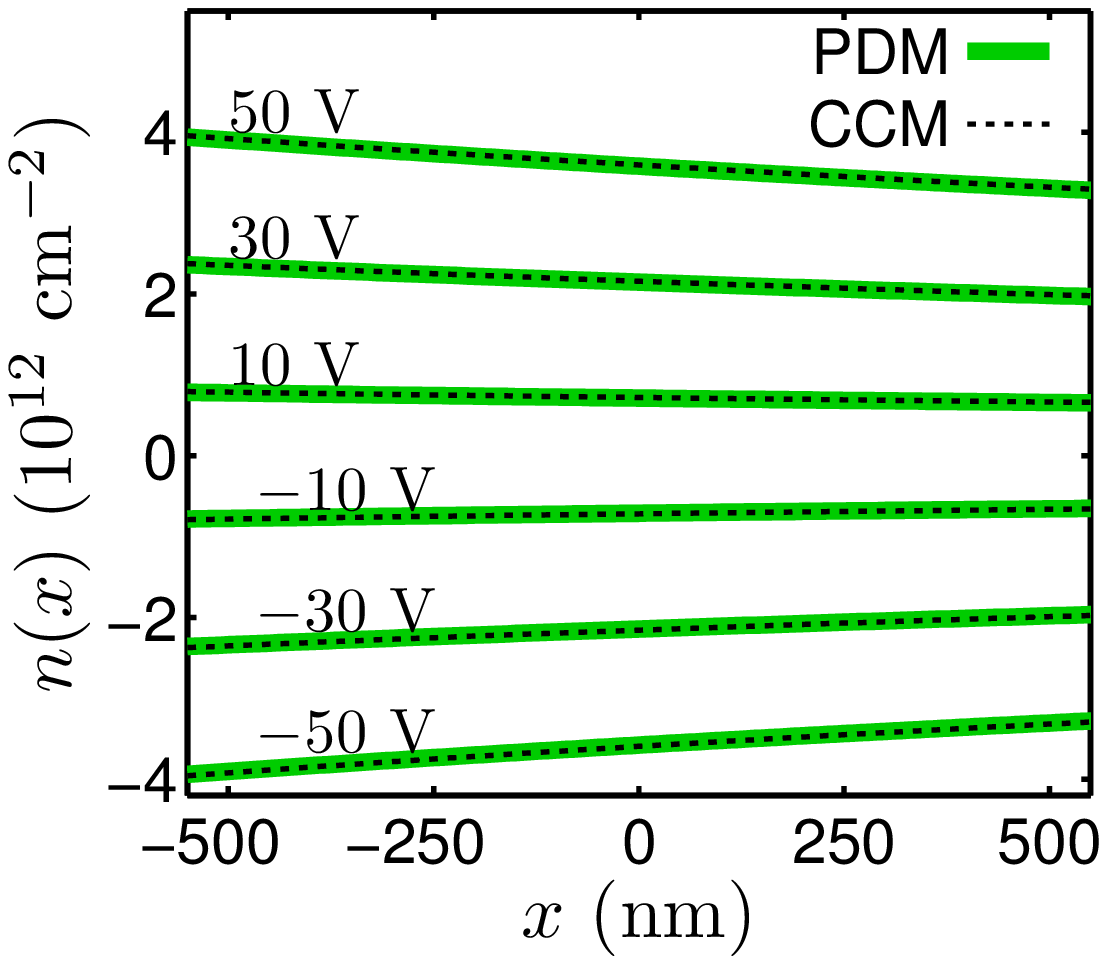}
\label{fig7b}} \hfill\subfigure[]{
\includegraphics[height=3.3cm]{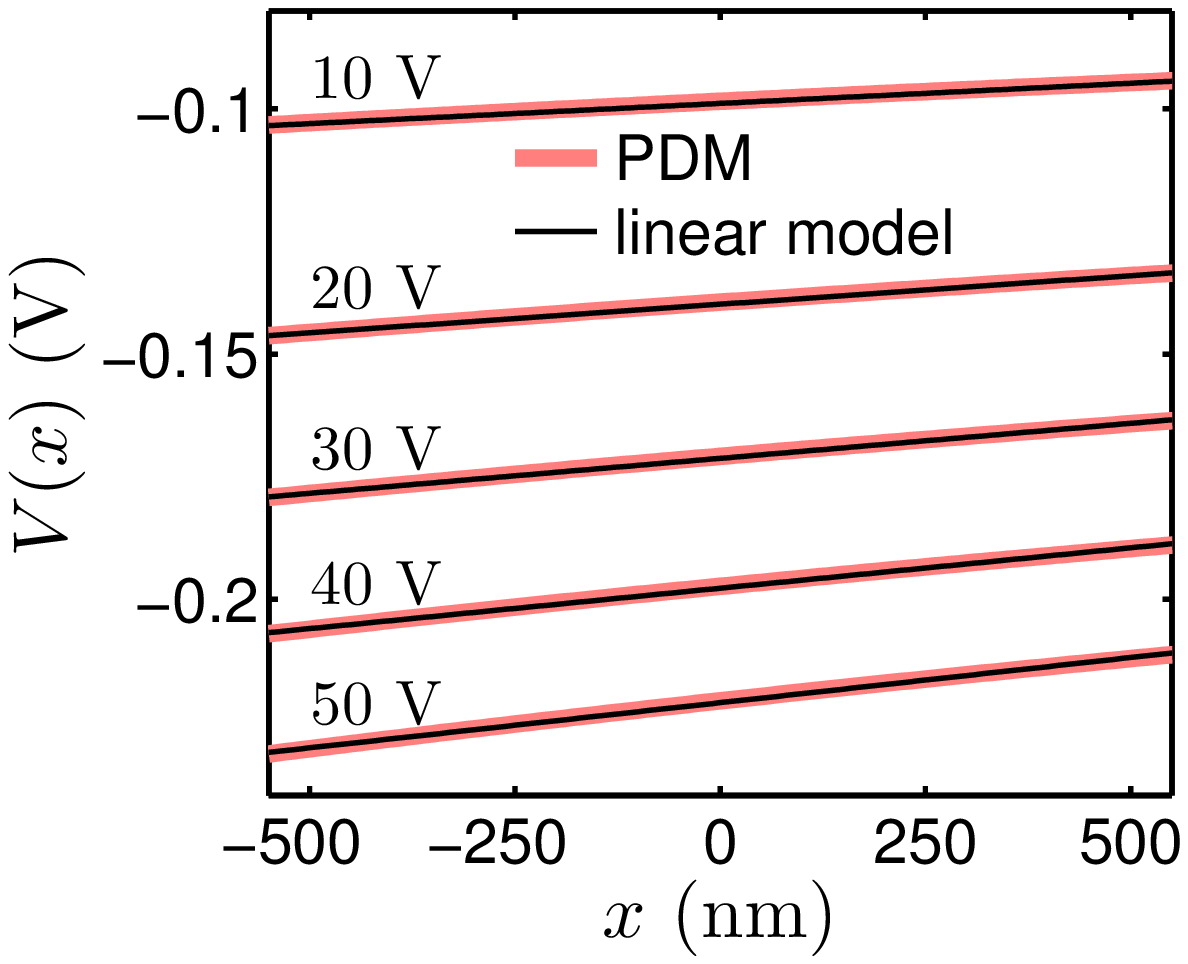}
\label{fig7c}}\caption{(a) Electrostatic potential $u(x,y)$ inside a
trapezoidal oxide layer. The tilted backgate with a slope of $0.05$ generates
a carrier density (b) that varies almost linearly with position. The
corresponding potential profiles (c) also exhibit a linear behavior. }%
\label{fig7}%
\end{figure}

In the previous example, we have demonstrated that fabricating a series of
patterned topgates may generate a periodic potential, which, combined with a
potential linear in position as well as the periodically alternating mass
potential, forms the prerequisite of the Bloch-Zener oscillation in graphene
\cite{Krueckl2012}. In this demonstrating example, we point out a simple way
to generate the linear potential: using a tilted backgate. As sketched in
Fig.\ \ref{fig7a}, where we consider a position-varying thickness of SiO$_{2}$
with a slope of $\allowbreak s=0.05$ (an increase of $50%
%TCIMACRO{\unit{nm}}%
%BeginExpansion
\operatorname{nm}%
%EndExpansion
$ per micron). In this case the quantum correction does not play a role (see
Sec.\ \ref{sec quantum correction}), and we show in Fig.\ \ref{fig7b} only the
carrier densities from the CCM and the PDM, which coincide to each other.

Since the classical capacitance model works well here, with the oxide
thickness $d(x)=d_{0}+sx$, where $d_{0}=300%
%TCIMACRO{\unit{nm}}%
%BeginExpansion
\operatorname{nm}%
%EndExpansion
$ is the thickness at the center, we can describe the carrier density as
$n(x)=\epsilon V_{g}/e(d_{0}+sx)$ and hence the potential as
$V(x)=-\operatorname*{sgn}(V_{g})\hbar v_{F}\sqrt{\pi\epsilon\left\vert
V_{g}\right\vert /e(d_{0}+sx)}$. The slope of the potential at $x=0$, $\left.
dV(x)/dx\right\vert _{x=0}$, together with the intercept $V(x=0)$, allows us
to approximate the potential with a linear model,%
\begin{equation}%
\begin{split}
V(x) &  \approx V_{0}+Sx\\
V_{0} &  =-\operatorname*{sgn}(V_{g})\sqrt{\frac{\epsilon_{r}\left\vert
V_{g}\right\vert }{d_{0}}}\times0.274\,25%
%TCIMACRO{\unit{eV}}%
%BeginExpansion
\operatorname{eV}%
%EndExpansion
\\
S &  =\operatorname*{sgn}(V_{g})\frac{s\sqrt{\epsilon_{r}\left\vert
V_{g}\right\vert }}{d_{0}^{3/2}}\times0.137\,13%
%TCIMACRO{\unit{eV}}%
%BeginExpansion
\operatorname{eV}%
%EndExpansion%
%TCIMACRO{\unit{nm}}%
%BeginExpansion
\operatorname{nm}%
%EndExpansion
^{-1}%
\end{split}
.\label{Vlinear}%
\end{equation}
In Fig.\ \ref{fig7c}, we plot the potential profiles obtained from
$n_{\text{PD}}$ and from the linear model given by Eq.\ \eqref{Vlinear}; the
consistency is almost perfect.

\subsection{Contact-induced doping and screening potential}

In the last application example, we turn to a practical issue for the graphene
electronics: contact-induced doping and its screening potential
\cite{Khomyakov2009,Khomyakov2010}. Taking palladium as the specific example,
we show that the solutions for the electric potential in graphene $V_{G}$
induced by the Pd contact solved by the PDM and by the QCM not only agree with
each other but also are consistent with the previously reported results based
on first principles \cite{Khomyakov2009,Khomyakov2010}, which we first briefly
review as follows.

Previous first-principles study expected ideal Pd(111) contacts to dope
graphene as \textit{n}-type, with the shift of the Fermi level given by
$\Delta E_{F}=W-W_{G}$, where $W$ is the work function of the metal coated
with graphene, and $W_{G}$ is the work function of free-standing graphene
\cite{Khomyakov2009}. In the case of Pd, $W=4.03%
%TCIMACRO{\unit{eV}}%
%BeginExpansion
\operatorname{eV}%
%EndExpansion
$, so $W_{G}=4.48%
%TCIMACRO{\unit{eV}}%
%BeginExpansion
\operatorname{eV}%
%EndExpansion
$ leads to $\Delta E_{F}=-0.45%
%TCIMACRO{\unit{eV}}%
%BeginExpansion
\operatorname{eV}%
%EndExpansion
$, according to the table summarized in \cite{Khomyakov2009}. This means that
the Fermi level of graphene coated by Pd(111) is expected to be $E_{F}=0.45%
%TCIMACRO{\unit{eV}}%
%BeginExpansion
\operatorname{eV}%
%EndExpansion
$. Furthermore, the contact-induced screening potential in graphene was
calculated in \cite{Khomyakov2010} using the density functional theory within
the Thomas-Fermi approximation, which is similar to the PDM introduced here
since the underlying equation that governs the electrostatic potential is
still the Poisson equation. In their formulation, the metal contact is
oriented at $x\leq0,z\geq0$ (graphene sheet also at $z=0)$, and the boundary
condition for the surface ($x=0,z\geq0$) of the contact is given by
$V_{c}=(W_{M}-W_{G})/e$, where $W_{M}$ is the work function of the
corresponding clean metal. In the case of Pd and again according to the table
summarized in \cite{Khomyakov2009}, $W_{M}=5.67%
%TCIMACRO{\unit{eV}}%
%BeginExpansion
\operatorname{eV}%
%EndExpansion
$, leading to $V_{c}=1.19%
%TCIMACRO{\unit{V}}%
%BeginExpansion
\operatorname{V}%
%EndExpansion
$.

The numerically exact solution for the electric potential in graphene, i.e.,
$V_{G}$ for $x\geq0$, was further fitted by a variational solution
\cite{Khomyakov2010}
\begin{equation}
V(x)\approx-\frac{V_{B}}{\left(  \sqrt{x/l_{s}+\beta_{2}^{2}}+\beta_{1}%
-\beta_{2}\right)  ^{1/2}\left(  x/l_{s}+\beta_{1}^{-2}\right)  ^{1/4}%
}\label{var}%
\end{equation}
with fitting parameters $\beta_{1}=0.915$ and $\beta_{2}=0.128$. The scaling
length $l_{s}$ in Eq.\ \eqref{var} is defined as $l_{s}=\hbar v_{F}/\pi
\alpha\left\vert V_{B}\right\vert $ with $\alpha=e^{2}/4\pi\epsilon
_{0}\epsilon_{r}\hbar v_{F}=2.\,\allowbreak187\,7/\epsilon_{r}$, where
$v_{F}=10^{8}%
%TCIMACRO{\unit{cm}}%
%BeginExpansion
\operatorname{cm}%
%EndExpansion
/%
%TCIMACRO{\unit{s}}%
%BeginExpansion
\operatorname{s}%
%EndExpansion
$ is assumed.\footnote{In \cite{Khomyakov2010}, $\alpha$ is given by
$2.38/\epsilon_{r}$ possibly because of the slightly different Fermi velocity
$v_{F}$.} For Pd, from their table (with $\beta=\pi/2$) one finds $V_{B}=0.48%
%TCIMACRO{\unit{eV}}%
%BeginExpansion
\operatorname{eV}%
%EndExpansion
$. Thus for vacuum with $\epsilon_{r}=1$, we have $l_{s}=0.199\,52%
%TCIMACRO{\unit{nm}}%
%BeginExpansion
\operatorname{nm}%
%EndExpansion
$. We will compare with this variational solution \eqref{var}.

\begin{figure}[t]
\includegraphics[width=\columnwidth]{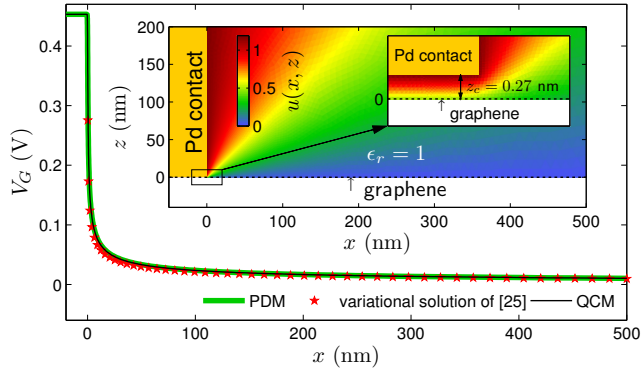}\caption{Contact-induced doping
and screening potential in the graphene sheet. The ideal Pd(111) contact is
modeled by a \textquotedblleft slightly floating gate\textquotedblright\ with
potential $V_{c}=(W_{M}-W_{G})/e=1.19\operatorname{V}$; quantities $W_{M}$ and
$W_{G}$ are described in the text. Inset: Schematic of the contact-graphene
system, with the electric potential $u(x,z)$ obtained by the PDM.}%
\label{fig8}%
\end{figure}

To apply the presently reviewed PDM and QCM to resolve the graphene electric
potential $V_{G}$, while keeping the results of
\cite{Khomyakov2009,Khomyakov2010} unchanged, namely,

\begin{enumerate}
\item $V_{G}(x\leq0)=E_{F}(x\leq0)/e=0.45%
%TCIMACRO{\unit{V}}%
%BeginExpansion
\operatorname{V}%
%EndExpansion
\equiv V_{G}^{<}$

\item $V_{G}(x\geq0)$ decays nonlinearly with $x$
\end{enumerate}

%

%TCIMACRO{\TeXButton{no indent}{\noindent}}%
%BeginExpansion
\noindent
%EndExpansion
we may model the metal contact as a slightly floating gate located at
$z=z_{c}$, as schematically shown in the inset of Fig.\ \ref{fig8}. By such
modeling, both of the above stated conclusions (i) and (ii) can be satisfied
at one time by taking the same input of the boundary condition. Specifically,
the boundary conditions to be applied here are $u(x=0,z\geq z_{c}%
)=u(x\leq0,z=z_{c})=V_{c}$, with the same $V_{c}=1.19%
%TCIMACRO{\unit{V}}%
%BeginExpansion
\operatorname{V}%
%EndExpansion
$ according to \cite{Khomyakov2010}.

To satisfy (i), we may apply the QCM for the parallel-plate capacitor to
deduce a proper floating height $z_{c}$ in order to meet the proper shift of
the Fermi level $V_{G}^{<}=E_{F}(x\leq0)/e=0.45%
%TCIMACRO{\unit{V}}%
%BeginExpansion
\operatorname{V}%
%EndExpansion
$. Thus the upper plate of the capacitor is the Pd contact with voltage
$V_{c}$, while the lower plate is graphene with voltage $V_{G}^{<}$. Equation
\eqref{VG by QCM} therefore reads
\begin{equation}
V_{G}^{<}=\frac{n_{Q}}{n_{C}}\left(  \sqrt{1+2\dfrac{\left\vert n_{C}%
\right\vert }{n_{Q}}}-1\right)  V_{c},\label{for zc}%
\end{equation}
where $n_{C}=C_{c}V_{c}/e$, with $C_{c}=\epsilon_{0}/z_{c}$ the classical
capacitance of the contact-vacuum-graphene sandwich, has been substituted.
Thus using $n_{C}/n_{Q}=(2V_{c}/\pi)(e/\hbar v_{F})^{2}(e/\epsilon_{0})z_{d}$
in Eq.\ \eqref{for zc}, one may solve for $z_{c}$ to obtain%
\begin{equation}
z_{c}=\pi\left(  \frac{\hbar v_{F}}{e}\right)  ^{2}\frac{\epsilon_{0}}{e}%
\frac{V_{c}-V_{G}^{<}}{(V_{G}^{<})^{2}}.\label{zc}%
\end{equation}
For the present case of Pd and following $V_{G}^{<}=0.45%
%TCIMACRO{\unit{V}}%
%BeginExpansion
\operatorname{V}%
%EndExpansion
$ of \cite{Khomyakov2009} and $V_{c}=1.19%
%TCIMACRO{\unit{V}}%
%BeginExpansion
\operatorname{V}%
%EndExpansion
$ of \cite{Khomyakov2010}, this effective height given by Eq.\ \eqref{zc}
amounts to $z_{c}\approx0.27%
%TCIMACRO{\unit{nm}}%
%BeginExpansion
\operatorname{nm}%
%EndExpansion
$.

Setting $z_{c}=0.27%
%TCIMACRO{\unit{nm}}%
%BeginExpansion
\operatorname{nm}%
%EndExpansion
$ for the contact and treating it as a \textquotedblleft
gate\textquotedblright\ with fixed potential $V_{c}=1.19%
%TCIMACRO{\unit{V}}%
%BeginExpansion
\operatorname{V}%
%EndExpansion
$, the electric potential in the graphene sheet $V_{G}$ is calculated by using
the self-consistent PDM and the analytical QCM, as shown in Fig.\ \ref{fig8}.
The two approaches again coincide with each other. In addition, $V_{G}%
(x\leq0)\approx0.45%
%TCIMACRO{\unit{V}}%
%BeginExpansion
\operatorname{V}%
%EndExpansion
$ is clearly observed, while the nonlinearly decaying $V_{G}(x\geq0)$ agrees
well with the variational solution of \cite{Khomyakov2010}, thus satisfying
both (i) and (ii).

Note that despite the consistency with the previous theory shown here,
experiments for transport measurements usually do not have single-crystal
contacts grown along (111), and the contact/graphene interface is certainly
dirty. The charge transfer between the metal contact and graphene due to their
different work functions is, therefore, greatly reduced, leading to a much
lower $V_{c}$. For example, a recent experiment observing the ballistic
interferences in ultra-clean suspended graphene uses Pd as contacts, and
theoretical modeling with $V_{c}$ of the order of $0.01%
%TCIMACRO{\unit{V}}%
%BeginExpansion
\operatorname{V}%
%EndExpansion
$ is found to better fit the transport measurement \cite{Rickhaus2013}.

\section{Summary\label{sec 5}}

In conclusion, theories of the gate-induced carrier density modulation in bulk
graphene have been reviewed. The classical capacitance model, the widely
adopted tool for carrier density estimation, does not include the quantum
correction but nevertheless plays usually the dominant role in the gate
modulation, unless the metal is rather close to graphene (Fig.\ \ref{fig2}).
The quantum correction stems from the finite capacity of the graphene sheet
for the electrons to reside, and can be treated by the self-consistent
Poisson-Dirac iteration method, as well as the exactly solvable quantum
capacitance model. By inspecting the numerical examples of single-gated
graphene, these two approaches are shown to agree with each other. In
particular, the correspondence is exact for the case of infinite
parallel-plate capacitors (Fig.\ \ref{fig3}). For the case with finite gates,
the agreement between QCM and PDM remains good (Fig.\ \ref{fig4}), implying
that Eq.\ \eqref{Cox(x)} is a good approximation for numerically calculating
the spatially varying capacitance. This further suggests that the classical
solution $n_{C}(x)$ serves as the preliminary solution step for the exactly
solvable QCM, circumventing the self-consistent iteration during the solution
process that is needed in the PDM. Along with the brief introduction to the
usage of the \textsc{Matlab} pdetool, the former part of this work
(Secs.\ \ref{sec 2}--\ref{sec 3}) provides a self-contained instruction to
calculating the carrier density of pristine graphene sheets subject to
complicated gating geometry. For the generalized theory for multigated doped
graphene, the readers are referred to \cite{Liu2013}.

To demonstrate the applicability of the introduced CCM, PDM, and QCM, the
latter part of this work (Sec.\ \ref{sec 4}) is devoted to illustration of
practical examples for calculating gate-induced carrier density in graphene
sheets, including the graphene \textit{pnp} junction using an embedded local
gate in addition to a global backgate (Fig.\ \ref{fig5}), graphene
superlattice potential by a series of patterned topgates (Fig.\ \ref{fig6}),
quasi-linear potential by using a tilted backgate (Fig.\ \ref{fig7}), and
finally the contact-induced doping and screening potential (Fig.\ \ref{fig8}).
The first three examples correspond to the experimental conditions that
provide a flexible platform to test the physics of Klein backscattering
\cite{Shytov2008,Young2009,Nam2011,Liu2012a} and the Bloch-Zener oscillation
\cite{Krueckl2012} in graphene, while the last example shows that the effects
of metal contacts can be treated equally well by the PDM and QCM, as compared
to the previous first-principles studies \cite{Khomyakov2009,Khomyakov2010}.
In either case, once the realistic potential profile $V(x)$ is obtained,
satisfactory electronic transport calculation for the relevant structure
following \cite{Liu2012a} can then be guaranteed.

\begin{acknowledgements}
The author thanks T.\ Fang and D.\ Jena for their illuminating suggestions,
F.-X.\ Schrettenbrunner, J.\ Eroms, P.\ Rickhaus, and R.\ Maurand for sharing their experimental viewpoints,
and V.\ Krueckl and K.\ Richter for valuable discussions.
Financial supports from Alexander von Humboldt Foundation (former part of the work)
and Deutsche Forschungsgemeinschaft within SFB 689 (present) are gratefully acknowledged.
\end{acknowledgements}


\end{document}